\documentclass[preprint2]{aastex}
\usepackage{spr-astr-addons}
\usepackage{url}
\usepackage{morefloats}
\bibpunct{(}{)}{,}{n}{,}{,}

\RequirePackage{color}

\begin{document}

   \title{Photometry of Centaurs and trans-Neptunian objects: 2060 Chiron (1977 UB), 
          10199 Chariklo (1997 CU$_{26}$), 38628 Huya (2000 EB$_{173}$), 28978 Ixion 
          (2001 KX$_{76}$), and 90482 Orcus (2004 DW)
         }
   \slugcomment{Based on observations collected at the Las Campanas Observatory.}

   \shorttitle{Photometry of Centaurs and TNOs}
   \shortauthors{Galiazzo et al.}

   \author{M.~Galiazzo}
   \affil{Department of Physics and Astronomy, The University of Western Ontario,
          London, ON N6A 3K7, Canada}
    \and
   \author{C.~de~la~Fuente Marcos}
    \and
   \author{R.~de~la~Fuente Marcos}
   \affil{Apartado de Correos 3413, E-28080 Madrid, Spain}
    \and
   \author{G. Carraro\altaffilmark{1}}
   \affil{European Southern Observatory, Alonso de Cordova 3107,
          Casilla 19001, Santiago 19, Chile}
    \and
   \author{M. Maris}
   \affil{INAF, Osservatorio Astronomico di Trieste,
          via G.B. Tiepolo 11, I-34131, Trieste, Italy}
    \and
   \author{M. Montalto}
   \affil{Centro de Astrof\'{\i}sica da Universidade do Porto, (CAUP),
        P-4150-762 Porto, Portugal}
   \email{gcarraro@eso.org}
   \altaffiltext{1}{Dipartimento di Fisica e Astronomia, Universit\`a degli Studi di Padova,
                    Vicolo dell'Osservatorio 3, I-35122, Padova, Italy.}

   \begin{abstract}
      Both Centaurs and trans-Neptunian objects (TNOs) are minor bodies found 
      in the outer Solar System. Centaurs are a transient population that 
      moves between the orbits of Jupiter and Neptune, and they probably 
      diffused out of the TNOs. TNOs move mainly beyond Neptune. Some of these 
      objects display episodic cometary behaviour; a few percent of them are 
      known to host binary companions. Here, we study the light-curves of two 
      Centaurs ---2060 Chiron (1977 UB) and 10199 Chariklo (1997 CU$_{26}$)--- 
      and three TNOs ---38628 Huya (2000 EB$_{173}$), 28978 Ixion (2001 
      KX$_{76}$), and 90482 Orcus (2004 DW)--- and the colours of the Centaurs 
      and Huya. Precise, $\sim$1\%, $R$-band absolute CCD photometry of these 
      minor bodies acquired between 2006 and 2011 is presented; the new data 
      are used to investigate the rotation rate of these objects. The colours 
      of the Centaurs and Huya are determined using $BVRI$ photometry. The 
      point spread function of the five minor bodies is analysed, searching 
      for signs of a coma or close companions. Astrometry is also discussed. A 
      periodogram analysis of the light-curves of these objects gives the 
      following rotational periods: 5.5$\pm$0.4~h for Chiron, 7.0$\pm$0.6~h 
      for Chariklo, 4.45$\pm$0.07~h for Huya, 12.4$\pm$0.3~h for Ixion, and 
      11.9$\pm$0.5~h for Orcus. The colour indices of Chiron are found to be 
      $B-V=0.53\pm0.05$, $V-R=0.37\pm0.08$, and $R-I=0.36\pm0.15$. The values 
      computed for Chariklo are $V-R=0.62\pm0.07$ and $R-I=0.61\pm0.07$. For 
      Huya, we find $V-R=0.58\pm0.09$ and $R-I=0.64\pm0.20$. Our rotation 
      periods are similar to and our colour values are consistent with those 
      already published for these objects. We find very low levels of cometary 
      activity (if any) and no sign of close or wide binary companions for 
      these minor bodies. 
   \end{abstract}


   \section{Introduction}
      Centaurs are a group of minor planets found in the outer Solar System whose orbits are strongly perturbed as a result of crossing the 
      paths of one or more of the giant planets (Di Sisto \& Brunini 2007; Galiazzo et al. 2015). Objects in this dynamical class are widely 
      thought to be former members of the so-called Trans-Neptunian Belt (TNB; e.g. Jewitt \& Luu 1993) or even the Oort Cloud (Levison et 
      al. 2001), and some of them may be transitioning to become short-period comets (e.g. Levison \& Duncan 1997). However, the possible 
      existence of trans-Plutonian planets (see e.g. Trujillo \& Sheppard 2014; de la Fuente Marcos \& de la Fuente Marcos 2014; de la
      Fuente Marcos et al. 2015; Batygin \& Brown 2016) may affect the dynamical pathways leading to this dynamical class. This transient 
      population features perihelia of less than 30~AU, but outside the orbit of Jupiter. A number of them display episodic cometary 
      behaviour (e.g. Jewitt 2009). Trans-Neptunian objects (TNOs) inhabit the TNB and their semi-major axes lie beyond that of Neptune.
      TNOs actively engaged in mean-motion resonances with Neptune are believed to have become trapped there during planet migration, late 
      in the giant-planet formation process (e.g. Gladman et al. 2012). Issues of nomenclature in the outer Solar System are discussed by 
      e.g. Gladman et al. (2008). A few percent of both Centaurs and TNOs are known to host binary companions (e.g. Walsh 2009; Naoz et al. 
      2010; Parker 2011; Parker et al. 2011).  

      In this paper, we present new photometric data of two Centaurs ---2060 Chiron (1977 UB) and 10199 Chariklo (1997 CU$_{26}$)--- and 
      three TNOs ---38628 Huya (2000 EB$_{173}$), 28978 Ixion (2001 KX$_{76}$), and 90482 Orcus (2004 DW). The observations are part of a 
      programme focused on the study of Centaurs, TNOs, and their possible cometary activity. Observations and data processing techniques 
      are described in Sect. 2. Chiron is revisited in Sect. 3. New data and results for Chariklo are presented in Sect. 4. Those for TNOs 
      Huya, Ixion, and Orcus are given in Sects. 5--7, respectively. Results are discussed and conclusions summarised in Sect. 8.

   \section{Observations and data reduction}
      The observations presented here were acquired at the Las Campanas Observatory, between 2006 and 2011, using the 1.0-m Swope telescope 
      equipped with the site$\#$3 $2048\times3150$ CCD camera. In the frames, the field of view is about $14\farcm8\times22\farcm8$ and the 
      pixel scale is 0.435\arcsec/pixel. Preliminary processing of the CCD frames was carried out using standard routines of the IRAF 
      package.\footnote{IRAF is distributed by the National Optical Astronomy Observatory, which is operated by the Association of 
      Universities for Research in Astronomy (AURA) under a cooperative agreement with the National Science Foundation.} Both dome and sky
      flat-field frames were obtained in each filter ($BVRI$) as needed and the images were also corrected for non-linearity (Hamuy et al. 
      2006; Carraro 2009); additional details can be found in Galiazzo (2009). Photometric calibration of the targets including appropriate 
      zero points and color terms computed for the individual observing nights ---when photometric--- was performed. As an example, the 
      following relationships between the instrumental (lower-case letters) and the standard colours and magnitudes were adopted in the 
      case of Chariklo (see Sect. 4): 
      \begin{eqnarray}
         V & = & 22.115(0.004) + v - 0.068(0.007) \times (B-V) \nonumber \\
           &   & + \ 0.16(0.02) \times X \,, \label{e1} \\
         B & = & 22.084(0.004) + b + 0.054(0.007) \times (B-V) \nonumber \\
           &   & + \ 0.30(0.02) \times X \,, \label{e2} \\
         I & = & 22.179(0.006) + i + 0.058(0.009) \times (V-I) \nonumber \\
           &   & + \ 0.06(0.02) \times X \,, \label{e3}
      \end{eqnarray}
      where $X$ is the airmass and the values of the errors associated with the various coefficients appear in parentheses. These 
      expressions were derived merging together standard stars from three different photometric nights after checking that their apparent 
      brightnesses were stable. Second order color terms were computed, but turned out to be negligible. Astrometric calibration of the CCD 
      frames was performed using the algorithms of the {\it Astrometry.net} system (Lang et al. 2010). In the data tables presented here, 
      when no magnitude is provided for a given astrometric entry (see Appendix A), it means that it was not computed because of the low 
      quality of the CCD frame or the presence of stars too close (even partially or completely blended with it) to the photometric target. 
      Due to the sparse nature of our data, period determination is made using one of the string-length period search algorithms, the 
      Lafler-Kinman method (Lafler \& Kinman 1965; Clarke 2002). String-length methods are better suited for this task when only a small 
      number of randomly spaced observations are available (e.g. Dworetsky 1983). False alarm probabilities are evaluated using the 
      Bootstrap method (e.g. Press et al. 2007) with 500 trials.

   \section{2060 Chiron (1977 UB)}
      Discovered in 1977, Chiron is the first known Centaur and one of the largest; its diameter amounts to 218$\pm$20~km, with an albedo of 
      16$\pm$3\% (Fornasier et al. 2013). The rotational period of Chiron as determined by Bus et al. (1989), Marcialis \& Buratti (1993) or
      Sheppard et al. (2008) amounts to 5.9178 h, but Fornasier et al. (2013) have found a value of 5.40$\pm$0.03~h in December 2011, with 
      an amplitude equal to 0.06--0.07 mag. This Centaur shows cometary activity (e.g. Luu \& Jewitt 1990) and it may have a ring (Ortiz et 
      al. 2015; Ruprecht et al. 2015). Because of this cometary activity, its absolute magnitude changes over time (Belskaya et al. 2010). 
      $BVRI$ photometry of Chiron was obtained in 2006 June and again in 2011 July--August (but only in $R$) using the same equipment. 

      During the first observing run (2006 June 26 to 30), Chiron was in Capricornus; as the object's apparent sky motion during the 
      observations was 0.11\arcsec/minute and the exposure time was 400 s, the expected shift during one exposure was well within the seeing 
      disk (1\farcs20). The average geocentric distance of Chiron during the observing run was $\bar{\Delta}$=13.32~AU, the average 
      heliocentric distance was $\bar{r}$=14.18~AU, and the average phase angle (Earth-Chiron-Sun angle) was $\bar{\alpha}$=2\fdg3. Table 
      \ref{photometrychironjune2006} includes the values of the apparent magnitude (Mag) and its associated error ($\pm\sigma$) at the 
      appropriate UT-time (Julian date), the filter used, the airmass (A.M.) and the solar phase angle $\alpha$ in degrees; the associated 
      astrometry is in Table \ref{astrometrychironjune2006}. In this and subsequent calculations the errors quoted correspond to one 
      standard deviation (1$\sigma$) computed applying the usual expressions (see e.g. Wall \& Jenkins 2012). A total of 25 frames obtained 
      in $R$ and one in each of the $B$, $V$, $R$, and $I$ filters are presented in Table \ref{photometrychironjune2006}. 

\begin{figure}
        \centering
        \includegraphics[width=\linewidth]{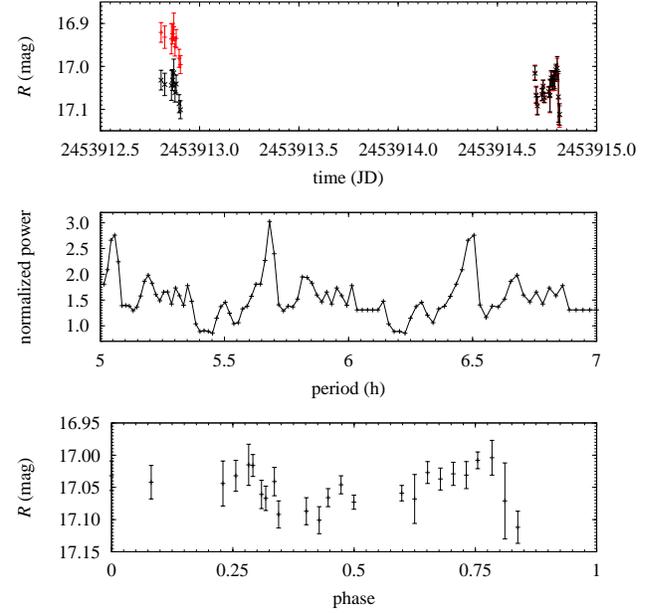}
        \caption{Chiron: 2006 June 26--30 run. $R$-band data, raw (red) and detrended (black), but uncorrected for rotational variation, 
                 used to compute the rotation period of Chiron (top panel). Lafler-Kinman periodogram of Chiron (middle panel) using 
                 4$\times$25 test frequencies. Periods yielding the lowest normalized power are the most likely, statistically speaking. The 
                 lowest value corresponds to a rotation period of 5.5$\pm$0.4 h or a frequency of 0.183$\pm$0.014 cycles/h with a false 
                 alarm probability (500 trials, see the text) of 1.6\%. Rotational light-curve of Chiron phased to a period of 5.5 hours 
                 (bottom panel).}
        \label{chi2006}
     \end{figure}

      The periodogram corresponding to the first run is shown in Fig. \ref{chi2006}, middle panel. Our best fitting gives a rotational 
      period $P=5.5\pm0.4$~h (or a frequency of 0.183$\pm$0.014 rotations per hour). The values of the false alarm probabilities are 
      relatively low: the probability that there is no period with value $P$ is 1.6$\pm$0.6\% and that of the observations containing a 
      period that is different from $P$ is $<$0.01\%. The light-curve of Chiron in Fig. \ref{chi2006}, bottom panel, shows the detrended 
      data (by fitting a linear function to the data and subtracting) from the top panel phased with the best-fit period, its amplitude is 
      $\sim$0.1 mag. Its light-curve amplitude was found to be 0.088 mag in 1986 and 1988 (Bus et al. 1989) and 0.044~mag in 1991 (Marcialis 
      \& Buratti 1993), but it was measured at 0.003$\pm$0.015~mag with data obtained in 2013 (Ortiz et al. 2015). The raw data in Fig. 
      \ref{chi2006} show a dimming trend that amounts to about 0.1~mag. Raw lightcurves often exhibit smooth trends with timescales of a few
      days. In our case (see also Figs. \ref{cha2006} and \ref{huy2006}), the smooth component seems to be roughly piecewise linear. There 
      are multiple reasons for this behaviour, including seeing-induced variability. It may also be intrinsic to the object under study. For
      example, fig. 6(a) in Luu \& Jewitt (1990) shows Chiron's rotational variations superposed on a linear brightening trend and this 
      could not be attributed to errors in the correction for atmospheric extinction; Chiron naturally exhibits short-term brightness 
      variations (on timescales of hours). Luu \& Jewitt (1990) removed the linear trend in their fig. 6(b) to facilitate their subsequent 
      rotational period analysis; the brightening trend amounted to about 0.12~mag which is consistent with our Fig. \ref{chi2006}, although 
      in our case we observe dimming not brightening. 

      Unfortunately, our sparse curve does not sample the entire rotational period well and this may explain why our value of $P$ is 
      somewhat smaller than the ones measured by other authors (i.e. Sheppard et al. 2008), although the accepted value is nearly within 
      1$\sigma$. However, it matches well the recent determination in Fornasier et al. (2013); the value of our amplitude is also consistent 
      with theirs. The value of the absolute magnitude in $V$ derived by Fornasier et al. (2013) is 5.80$\pm$0.04 mag; ours is 5.75$\pm$0.06 
      mag which again is compatible with theirs and also with the mean value found over 2004--2008 (see fig. 2 in Fornasier et al. 2013). 
      The absolute magnitude in $V$ has been computed using eqs. (1) and (2) in Romanishin \& Tegler (2005).

      Our photometric data are compatible with a negligible level of cometary activity at the time of the observations and we found no 
      evidence for a comoving (close or wide) companion (see Sect. 8 for additional details). Regarding the colours of this object, we only 
      used consecutive images, or almost consecutive, on the $BVRI$ filters. Adopting this approach we avoid errors induced by possible 
      rotational variability associated with surface features. The colour indices of Chiron were found to be $B-V=0.53\pm0.05$, 
      $V-R=0.37\pm0.08$, and $R-I=0.36\pm0.15$. Our values are consistent with those already published for this object (e.g. Hainaut \& 
      Delsanti 2002; Barucci et al. 2005). 
%
%
     \begin{table}
        \fontsize{8}{11pt}\selectfont
        \tabcolsep 0.10truecm
        \caption{Photometry of 2060 Chiron (1977 UB); 2006 June. This table includes the values of the apparent magnitude (Mag) and its 
                 associated error ($\pm\sigma$) at the appropriate UT-time (Julian date), the filter used, the airmass (A.M.) and the solar 
                 phase angle $\alpha$ in degrees.}
        \centering
        \begin{tabular}{lccccc}
           \hline
           \hline
            \multicolumn{1}{c}{Julian date}            &
            \multicolumn{1}{c}{Filter}                 &
            \multicolumn{1}{c}{A.M.}                   &
            \multicolumn{1}{c}{Exp. (s)}               &
            \multicolumn{1}{c}{Mag}                    &
            \multicolumn{1}{c}{$\alpha$ (\degr)}       \\
           \hline
            2453912.805822 & $R$ & 1.05 & 400 & 16.921$\pm$0.023 & 2.36 \\
            2453912.824410 & $R$ & 1.07 & 400 & 16.932$\pm$0.026 & 2.36 \\
            2453912.857975 & $R$ & 1.13 & 400 & 16.936$\pm$0.035 & 2.35 \\
            2453912.864005 & $R$ & 1.15 & 400 & 16.924$\pm$0.024 & 2.35 \\
            2453912.870023 & $R$ & 1.16 & 400 & 16.908$\pm$0.032 & 2.35 \\
            2453912.876053 & $R$ & 1.19 & 400 & 16.954$\pm$0.022 & 2.35 \\
            2453912.882083 & $R$ & 1.21 & 400 & 16.935$\pm$0.021 & 2.35 \\
            2453912.897037 & $R$ & 1.28 & 400 & 16.981$\pm$0.021 & 2.35 \\
            2453912.903067 & $R$ & 1.31 & 400 & 16.996$\pm$0.021 & 2.35 \\
            2453914.689248 & $R$ & 1.28 & 400 & 17.014$\pm$0.017 & 2.25 \\
            2453914.695278 & $R$ & 1.25 & 400 & 17.065$\pm$0.019 & 2.25 \\
            2453914.701296 & $R$ & 1.22 & 400 & 17.091$\pm$0.021 & 2.25 \\
            2453914.724491 & $R$ & 1.14 & 400 & 17.066$\pm$0.014 & 2.25 \\
            2453914.730521 & $R$ & 1.13 & 400 & 17.046$\pm$0.014 & 2.25 \\
            2453914.736539 & $R$ & 1.11 & 400 & 17.074$\pm$0.011 & 2.25 \\
            2453914.758970 & $R$ & 1.07 & 400 & 17.061$\pm$0.012 & 2.24 \\
            2453914.765000 & $R$ & 1.07 & 400 & 17.070$\pm$0.038 & 2.24 \\
            2453914.771030 & $R$ & 1.06 & 400 & 17.030$\pm$0.017 & 2.24 \\
            2453914.777049 & $R$ & 1.06 & 400 & 17.040$\pm$0.017 & 2.24 \\
            2453914.783079 & $R$ & 1.05 & 400 & 17.032$\pm$0.018 & 2.24 \\
            2453914.789167 & $R$ & 1.05 & 400 & 17.035$\pm$0.021 & 2.24 \\
            2453914.794502 & $R$ & 1.05 & 400 & 17.012$\pm$0.013 & 2.24 \\
            2453914.801227 & $R$ & 1.05 & 400 & 17.008$\pm$0.027 & 2.24 \\
            2453914.807245 & $R$ & 1.06 & 400 & 17.076$\pm$0.059 & 2.24 \\
            2453914.813275 & $R$ & 1.06 & 400 & 17.117$\pm$0.025 & 2.24 \\
           \hline
            2453912.798391 & $B$ & 1.05 & 600 & 17.815$\pm$0.025 & 2.36 \\
           \hline
            2453912.811991 & $V$ & 1.06 & 400 & 17.287$\pm$0.026 & 2.36 \\
           \hline
            2453912.818206 & $I$ & 1.06 & 400 & 16.564$\pm$0.590 & 2.36 \\
           \hline
        \end{tabular}
        \label{photometrychironjune2006}
     \end{table}
%
%
%
%
%
%

      Chiron was reobserved from 2011 July 30 to August 6 (see Tables \ref{photometrychiron2011} and \ref{astrometrychiron2011}). The 
      average geocentric distance of Chiron during this second observing run was $\bar{\Delta}$=15.92~AU, the average heliocentric distance 
      was $\bar{r}$=16.84~AU, and the average phase angle was $\bar{\alpha}$=1\fdg5. Unfortunately, the nights were not photometric and the 
      seeing was variable and worse than that of the first run. Apparent magnitudes in Table \ref{photometrychiron2011} are relative to two
      suitable reference field stars (not known variables) close to the target body. An additional, exploratory observing run for Chiron was 
      carried out with the Cerro Tololo InterAmerican Observatory SMARTS 1-m telescope and Y4kCam on 2006 May 19 to 22 (see Table 
      \ref{astrometrychironmay2006}). During this initial observing run only astrometry was obtained. The average geocentric distance of 
      Chiron was $\bar{\Delta}$=13.74~AU, the average heliocentric distance was $\bar{r}$=14.11~AU, and the average phase angle was 
      $\bar{\alpha}$=3\fdg9.
%
%
     \begin{table}
        \fontsize{8}{11pt}\selectfont
        \tabcolsep 0.20truecm
        \caption{Photometry of 2060 Chiron (1977 UB); 2011 July--August. All the observations in the $R$ filter; notation as in Table 
                 \ref{photometrychironjune2006}.}
        \centering
        \begin{tabular}{lcccc}
           \hline
           \hline
            \multicolumn{1}{c}{Julian date}            &
            \multicolumn{1}{c}{A.M.}                   &
            \multicolumn{1}{c}{Exp. (s)}               &
            \multicolumn{1}{c}{Mag}                    &
            \multicolumn{1}{c}{$\alpha$ (\degr)}       \\
           \hline
            2455772.717766 & 1.17 & 600 & 19.557$\pm$0.075 & 1.64 \\
            2455772.726655 & 1.15 & 600 & 19.370$\pm$0.061 & 1.64 \\
            2455772.738704 & 1.13 & 800 & 18.210$\pm$0.027 & 1.64 \\
            2455772.750451 & 1.12 & 800 & 17.233$\pm$0.011 & 1.64 \\
            2455772.761412 & 1.11 & 800 & 17.325$\pm$0.012 & 1.64 \\
            2455772.772199 & 1.10 & 800 & 16.899$\pm$0.008 & 1.64 \\
            2455772.783102 & 1.11 & 800 & 16.864$\pm$0.007 & 1.64 \\
            2455772.793854 & 1.11 & 800 & 16.864$\pm$0.008 & 1.63 \\
            2455772.804618 & 1.12 & 800 & 16.875$\pm$0.008 & 1.63 \\
            2455772.815347 & 1.14 & 800 & 16.866$\pm$0.008 & 1.63 \\
            2455772.829167 & 1.17 & 800 & 16.869$\pm$0.008 & 1.63 \\
            2455772.840347 & 1.20 & 800 & 17.156$\pm$0.009 & 1.63 \\
            2455772.850046 & 1.24 & 600 & 17.347$\pm$0.011 & 1.63 \\
            2455772.858657 & 1.27 & 600 & 17.202$\pm$0.010 & 1.63 \\
            2455772.867060 & 1.32 & 600 & 17.197$\pm$0.010 & 1.63 \\
            2455772.875475 & 1.36 & 600 & 17.188$\pm$0.010 & 1.63 \\
            2455772.883889 & 1.42 & 600 & 17.190$\pm$0.010 & 1.63 \\
            2455772.892292 & 1.48 & 600 & 17.191$\pm$0.010 & 1.63 \\
            2455772.900706 & 1.56 & 600 & 17.227$\pm$0.010 & 1.63 \\
            2455772.909144 & 1.64 & 600 & 17.202$\pm$0.011 & 1.63 \\
            2455775.694271 & 1.22 & 600 & 17.714$\pm$0.015 & 1.48 \\
            2455775.702766 & 1.19 & 600 & 19.942$\pm$0.117 & 1.48 \\
            2455775.712813 & 1.16 & 600 & 19.013$\pm$0.049 & 1.48 \\
            2455775.723380 & 1.14 & 900 & 19.542$\pm$0.097 & 1.48 \\
            2455775.746100 & 1.11 & 900 & 18.439$\pm$0.034 & 1.48 \\
            2455775.758449 & 1.11 & 900 & 17.691$\pm$0.020 & 1.48 \\
            2455775.770694 & 1.10 & 900 & 16.889$\pm$0.008 & 1.48 \\
            2455775.792616 & 1.12 & 900 & 16.938$\pm$0.009 & 1.48 \\
            2455775.805799 & 1.14 & 900 & 16.718$\pm$0.007 & 1.48 \\
            2455775.816470 & 1.16 & 600 & 17.972$\pm$0.020 & 1.48 \\
            2455775.860116 & 1.33 & 600 & 17.992$\pm$0.020 & 1.47 \\
            2455775.871898 & 1.39 & 600 & 17.700$\pm$0.016 & 1.47 \\
            2455775.880336 & 1.46 & 600 & 17.546$\pm$0.013 & 1.47 \\
            2455775.888762 & 1.53 & 600 & 17.591$\pm$0.015 & 1.47 \\
            2455775.897187 & 1.61 & 600 & 17.265$\pm$0.011 & 1.47 \\
            2455775.905613 & 1.70 & 600 & 17.464$\pm$0.014 & 1.47 \\
            2455775.914039 & 1.82 & 600 & 17.210$\pm$0.011 & 1.47 \\
            2455779.679375 & 1.23 & 600 & 17.099$\pm$0.011 & 1.27 \\
            2455779.687778 & 1.20 & 600 & 17.109$\pm$0.011 & 1.27 \\
            2455779.697523 & 1.17 & 600 & 17.076$\pm$0.010 & 1.27 \\
            2455779.705926 & 1.15 & 600 & 17.033$\pm$0.010 & 1.27 \\
            2455779.714317 & 1.14 & 600 & 16.623$\pm$0.006 & 1.27 \\
            2455779.722847 & 1.12 & 600 & 16.808$\pm$0.008 & 1.27 \\
            2455779.731238 & 1.11 & 600 & 17.138$\pm$0.028 & 1.27 \\
            2455779.765116 & 1.11 & 600 & 17.047$\pm$0.059 & 1.26 \\
            2455779.773600 & 1.11 & 600 & 16.797$\pm$0.007 & 1.26 \\
            2455779.781991 & 1.12 & 600 & 17.099$\pm$0.011 & 1.26 \\
            2455779.790394 & 1.13 & 600 & 17.092$\pm$0.009 & 1.26 \\
            2455779.814792 & 1.18 & 900 & 16.856$\pm$0.010 & 1.26 \\
            2455779.825417 & 1.22 & 600 & 17.123$\pm$0.010 & 1.26 \\
            2455779.833854 & 1.25 & 600 & 17.108$\pm$0.010 & 1.26 \\
            2455779.842245 & 1.29 & 600 & 17.099$\pm$0.009 & 1.26 \\
            2455779.850694 & 1.34 & 600 & 17.134$\pm$0.010 & 1.26 \\
            2455779.859097 & 1.39 & 600 & 17.231$\pm$0.010 & 1.26 \\
            2455779.867488 & 1.44 & 600 & 17.138$\pm$0.009 & 1.26 \\
            2455779.876146 & 1.51 & 600 & 17.113$\pm$0.009 & 1.26 \\
            2455779.884560 & 1.59 & 600 & 17.140$\pm$0.009 & 1.26 \\
            2455779.893032 & 1.69 & 600 & 17.144$\pm$0.010 & 1.26 \\
            2455779.901435 & 1.80 & 600 & 17.133$\pm$0.010 & 1.26 \\
           \hline
        \end{tabular}
        \label{photometrychiron2011}
     \end{table}
%
%

   \section{10199 Chariklo (1997 CU$_{26}$)}
      Currently the largest confirmed Centaur, Chariklo has a rotational period of 7.00$\pm$0.04 h, an effective radius of 119$\pm$5~km 
      (Fornasier et al. 2014), and a dense ring system (Braga-Ribas et al. 2014; Duffard et al. 2014). Due to the rings, the average value 
      of the albedo, 4.2\% (Fornasier et al. 2014), is variable (Duffard et al. 2014). $VRI$ photometry of Chariklo was obtained in 2006 
      June and again in 2011 July--August using the same equipment. During the first observing run, Chariklo was in Hydra and the 
      observations were completed during 5 nights, 3 (2006 June 26, 28 and 29) of them photometric. During the second run, that lasted 6 
      nights, Chariklo was in Lupus. These nights were not photometric and only frames in the $R$-band were collected. 

\begin{figure}
        \centering
        \includegraphics[width=\linewidth]{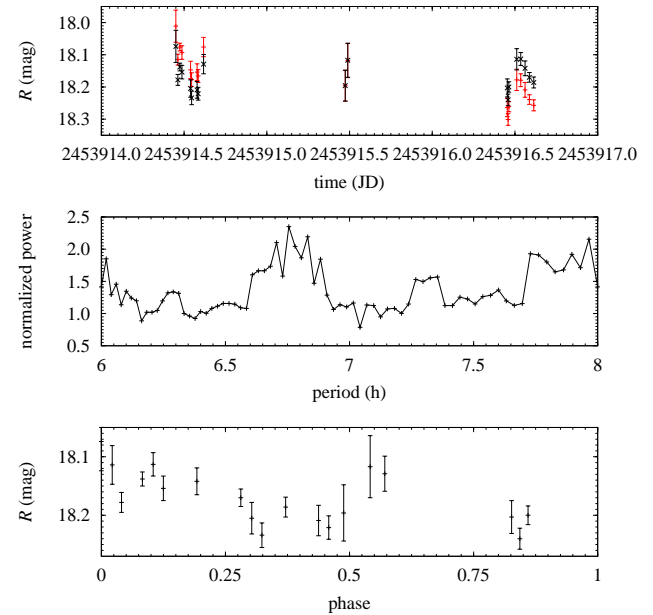}
        \caption{Same as Fig. \ref{chi2006} but for Chariklo; 4$\times$19 test frequencies. The lowest value of the normalized power 
                 corresponds to a rotation period of 7.0$\pm$0.6 h or a frequency of 0.142$\pm$0.013 cycles/h with a false alarm 
                 probability $<$ 4.5\%. The bottom panel shows the rotational light-curve of Chariklo phased to a period of 7.0 hours.}
        \label{cha2006}
     \end{figure}

      For the first observing run, preliminary processing of the CCD frames was made as described in Sect. 2. The average seeing was 
      1\farcs20 along the entire 5 nights run (2006 June 26 to 30). The sky motion of Chariklo during this first run was 0.071\arcsec/minute 
      and the typical exposure time was 400 s; therefore, the expected shift within a given image is again well within the seeing disk. The 
      average geocentric distance of Chariklo during the observing run was $\bar{\Delta}$ = 12.98 AU, the average heliocentric distance was 
      $\bar{r}$=13.20~AU, and the average phase angle (Earth--Chariklo--Sun angle) was $\bar{\alpha}$ = 4\fdg35. On each of the photometric 
      nights, we observed 96 standard stars extracted from repeated observations of the four fields Mark A, PG 1657, PG 2213 and SA 110 
      (Landolt 1992) at different airmasses. Aperture photometry of standard stars was obtained with an aperture radius of 6$\farcs$09 (14 
      pixels). The instrumental photometry of Chariklo and several field stars was extracted with the DAOPHOTII (Stetson 1987) package, with 
      a fitting radius of 5 pixels. We used five stable field stars as reference to shift Chariklo's magnitudes to the first night (June 
      27th), which was photometric. We then estimated aperture corrections for the field stars, that we applied to Chariklo's measurements 
      ---see Carraro et al. (2006) and Galiazzo (2009) for further details. This correction turned out to be smaller than 0.10 mag in all 
      filters. Table \ref{run06} is analogous in structure to Table \ref{photometrychironjune2006} and includes Chariklo's data details for 
      the 2006 run; the associated astrometry can be found in Table \ref{astrometry2006}. The data are plotted in Fig. \ref{cha2006}, top 
      panel. 

      The periodogram for the first run (Fig. \ref{cha2006}, middle panel) shows a broad minimum between 6.8 and 7.3 h. Our best fit for the 
      rotational period is $P=7.0\pm0.6$~h (or a frequency of 0.142$\pm$0.013 rotations per hour). The probability that there is no period 
      with value $P$ is 4.2$\pm$0.9\% and that of the observations containing a period that is different from $P$ is 0.2$\pm$0.2\%. The 
      light-curve of Chariklo in Fig. \ref{cha2006}, bottom panel, represents the detrended data from the top panel phased with the best-fit 
      period. Our sparse curve matches well that in fig. 1 of Fornasier et al. (2014), which has an amplitude of 0.11 mag: it displays 
      asymmetric double peaks and an amplitude of $\sim$0.13 mag. The amplitude in Lacerda \& Luu (2006) is about 0.1 mag. The rings affect 
      the amplitude of the overall rotational light-curve as their aspect angle changes over time (Duffard et al. 2014). The value of the 
      absolute magnitude in $V$ found by Fornasier et al. (2014) is 7.03$\pm$0.10 mag; our determination (found as described in Sect. 3), 
      7.24$\pm$0.08 mag, is compatible with this value. Consistently with results in Fornasier et al. (2014) and Duffard et al. (2014) no 
      coma was detected (but see Sect. 8 for additional details).

      Regarding the colour indices, the values found for Chariklo were $B-V=0.80\pm0.05$, $V-R=0.62\pm0.07$, and $R-I=0.61\pm0.07$. Again, 
      the values are consistent with some already published for this object (e.g. Hainaut \& Delsanti 2002). However, they are different 
      from those in Fulchignoni et al. (2008): $V-R=0.48$, $V-I=1.01$. These significant differences may be the result of changes in the 
      appearance of the rings that induce variations in the overall spectral properties of this object as the absorption band due to
      water ice disappears when the rings are edge-on (Duffard et al. 2014; Fornasier et al. 2014).  
%
%
     \begin{table}
        \fontsize{8}{11pt}\selectfont
        \tabcolsep 0.10truecm
        \caption{Photometry of 10199 Chariklo (1997 CU$_{26}$); 2006 June. Notation as in Table \ref{photometrychironjune2006}.}
        \centering
        \begin{tabular}{lccccc}
           \hline
           \hline
            \multicolumn{1}{c}{Julian date}      &
            \multicolumn{1}{c}{Filter}           &
            \multicolumn{1}{c}{A.M.}             &
            \multicolumn{1}{c}{Exp. (s)}         &
            \multicolumn{1}{c}{Mag}              &
            \multicolumn{1}{c}{$\alpha$ (\degr)} \\
           \hline
            2453914.450405 & $R$ & 1.00 & 300 & 18.011$\pm$0.050 & 4.34 \\
            2453914.462315 & $R$ & 1.00 & 400 & 18.115$\pm$0.017 & 4.34 \\
            2453914.474676 & $R$ & 1.01 & 400 & 18.076$\pm$0.012 & 4.34 \\
            2453914.487072 & $R$ & 1.02 & 400 & 18.093$\pm$0.021 & 4.34 \\
            2453914.539294 & $R$ & 1.14 & 400 & 18.147$\pm$0.027 & 4.34 \\
            2453914.545313 & $R$ & 1.16 & 400 & 18.177$\pm$0.021 & 4.34 \\
            2453914.578819 & $R$ & 1.32 & 400 & 18.154$\pm$0.026 & 4.34 \\
            2453914.584850 & $R$ & 1.36 & 400 & 18.166$\pm$0.020 & 4.34 \\
            2453914.617986 & $R$ & 1.65 & 400 & 18.076$\pm$0.030 & 4.34 \\
            2453915.473970 & $R$ & 1.01 & 400 & 18.196$\pm$0.048 & 4.35 \\
            2453915.489641 & $R$ & 1.03 & 400 & 18.118$\pm$0.053 & 4.35 \\
            2453916.453565 & $R$ & 1.00 & 400 & 18.264$\pm$0.028 & 4.36 \\
            2453916.458426 & $R$ & 1.00 & 400 & 18.301$\pm$0.018 & 4.36 \\
            2453916.463299 & $R$ & 1.01 & 400 & 18.261$\pm$0.016 & 4.36 \\
            2453916.510891 & $R$ & 1.07 & 400 & 18.178$\pm$0.033 & 4.36 \\
            2453916.535081 & $R$ & 1.14 & 400 & 18.179$\pm$0.020 & 4.36 \\
            2453916.560914 & $R$ & 1.25 & 400 & 18.209$\pm$0.023 & 4.37 \\
            2453916.586921 & $R$ & 1.40 & 500 & 18.239$\pm$0.015 & 4.37 \\
            2453916.613403 & $R$ & 1.65 & 500 & 18.257$\pm$0.017 & 4.37 \\
           \hline
            2453914.456100 & $V$ & 1.00 & 400 & 18.677$\pm$0.024 & 4.34 \\
            2453914.468484 & $V$ & 1.01 & 400 & 18.709$\pm$0.027 & 4.34 \\
           \hline
            2453914.480833 & $I$ & 1.01 & 400 & 17.542$\pm$0.018 & 4.34 \\
           \hline
        \end{tabular}
        \label{run06}
     \end{table}
%
%
%
%
%
%

      Conditions during the second run were less favourable with the seeing changing in the range 1\farcs2--2\farcs0. No standard stars were 
      used in this case. As the nights were not photometric, no attempt is made to calibrate with respect to standards (see Tables 
      \ref{run11} and \ref{astrometry2011}). Apparent magnitudes in Table \ref{run11} are relative to two suitable reference field stars 
      (not known variables) close to the target body. The average geocentric distance of Chariklo during this second observing run was 
      $\bar{\Delta}$=13.72~AU, the average heliocentric distance was $\bar{r}$=14.08~AU, and the average phase angle was 
      $\bar{\alpha}$=3\fdg9.
%
%
     \begin{table}
        \fontsize{8}{11pt}\selectfont
        \tabcolsep 0.20truecm
        \caption{Photometry of 10199 Chariklo (1997 CU$_{26}$); 2011 July-August. All the observations in the $R$ filter.}
        \centering
        \begin{tabular}{lcccc}
           \hline
           \hline
            \multicolumn{1}{c}{Julian date}      &
            \multicolumn{1}{c}{A.M.}             &
            \multicolumn{1}{c}{Exp. (s)}         &
            \multicolumn{1}{c}{Mag}              &
            \multicolumn{1}{c}{$\alpha$ (\degr)} \\
           \hline
            2455772.543628 & 1.05 & 600 & 18.317$\pm$0.013 & 3.83 \\
            2455772.584763 & 1.14 & 800 & 18.523$\pm$0.232 & 3.83 \\
            2455772.600174 & 1.20 & 800 & 18.321$\pm$0.047 & 3.83 \\
            2455772.611632 & 1.24 & 800 & 18.282$\pm$0.029 & 3.83 \\
            2455772.622697 & 1.29 & 800 & 18.243$\pm$0.040 & 3.83 \\
            2455772.634063 & 1.36 & 800 & 18.284$\pm$0.030 & 3.83 \\
            2455772.645208 & 1.43 & 800 & 18.312$\pm$0.054 & 3.83 \\
            2455772.656111 & 1.51 & 800 & 18.518$\pm$0.138 & 3.83 \\
            2455775.527824 & 1.04 & 600 & 18.286$\pm$0.018 & 3.90 \\
            2455775.536250 & 1.06 & 600 & 18.249$\pm$0.037 & 3.90 \\
            2455775.553137 & 1.09 & 600 & 18.259$\pm$0.021 & 3.90 \\
            2455775.561562 & 1.11 & 600 & 18.246$\pm$0.020 & 3.90 \\
            2455775.569988 & 1.13 & 600 & 18.213$\pm$0.018 & 3.90 \\
            2455775.578426 & 1.15 & 600 & 18.215$\pm$0.024 & 3.90 \\
            2455775.586863 & 1.18 & 600 & 18.225$\pm$0.023 & 3.90 \\
            2455775.595289 & 1.21 & 600 & 18.240$\pm$0.020 & 3.90 \\
            2455775.603727 & 1.25 & 600 & 18.210$\pm$0.020 & 3.90 \\
            2455775.616771 & 1.31 & 600 & 18.185$\pm$0.021 & 3.90 \\
            2455775.626400 & 1.37 & 600 & 18.205$\pm$0.020 & 3.90 \\
            2455775.670139 & 1.74 & 600 & 18.267$\pm$0.019 & 3.90 \\
            2455775.678553 & 1.85 & 600 & 18.249$\pm$0.020 & 3.90 \\
            2455778.509792 & 1.03 & 600 & 18.234$\pm$0.026 & 3.95 \\
            2455778.518194 & 1.04 & 600 & 18.264$\pm$0.028 & 3.95 \\
            2455778.526620 & 1.05 & 600 & 18.256$\pm$0.023 & 3.95 \\
            2455778.536684 & 1.07 & 600 & 18.266$\pm$0.029 & 3.95 \\
            2455778.545590 & 1.09 & 600 & 18.262$\pm$0.022 & 3.95 \\
            2455778.573171 & 1.16 & 600 & 18.301$\pm$0.019 & 3.95 \\
            2455778.581586 & 1.19 & 600 & 18.300$\pm$0.024 & 3.95 \\
            2455778.619340 & 1.37 & 600 & 18.307$\pm$0.020 & 3.95 \\
            2455778.627755 & 1.43 & 600 & 18.277$\pm$0.020 & 3.95 \\
            2455778.636169 & 1.49 & 600 & 18.270$\pm$0.018 & 3.95 \\
            2455778.644676 & 1.57 & 600 & 18.226$\pm$0.019 & 3.95 \\
            2455778.653090 & 1.65 & 600 & 18.226$\pm$0.020 & 3.95 \\
           \hline
        \end{tabular}
        \label{run11}
     \end{table}
%
%
  
   \section{38628 Huya (2000 EB$_{173}$)}
\begin{figure}
       \centering
        \includegraphics[width=\linewidth]{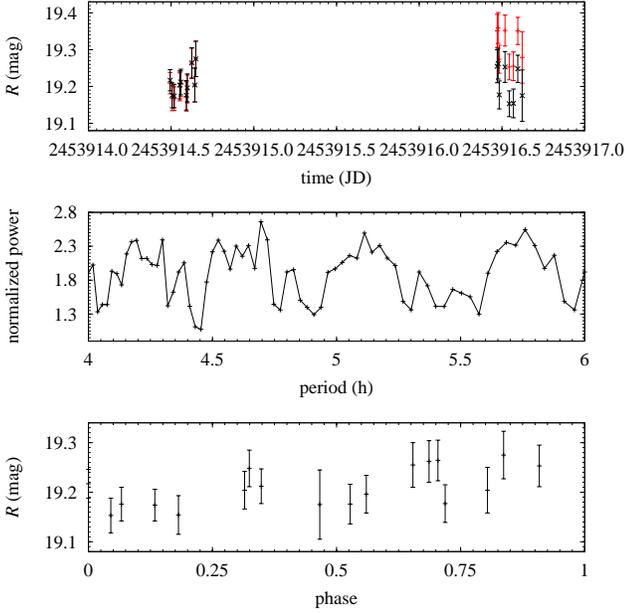}
        \caption{Same as Fig. \ref{chi2006} but for Huya; 4$\times$18 test frequencies. The lowest value of the normalized power corresponds 
                 to a rotation period of 4.45$\pm$0.07 h or a frequency of 0.225$\pm$0.003 cycles/h with a false alarm probability $<$ 
                 33\%. The bottom panel shows the rotational light-curve of Huya phased to a period of 4.45 h.
                }
        \label{huy2006}
     \end{figure}

      Huya is a TNO trapped in a 3:2 mean motion resonance with Neptune, it is therefore a Plutino. Its diameter could be as large as 
      458$\pm$9~km, with an albedo of 8.3$\pm$0.4\% (Fornasier et al. 2013). Thirouin et al. (2014) give a value of 5.28 h for the 
      rotational period of this object; Ortiz et al. (2003) derived a value of 6.75 h. Huya was observed with Chiron and Chariklo in 2006 
      June (see Tables \ref{huya} and \ref{astrometryhuya2006}); it was in Virgo. The average geocentric distance of Huya during the 
      observing run was $\bar{\Delta}$ = 28.59 AU, the average heliocentric distance was $\bar{r}$=29.03~AU, and the average phase angle 
      (Earth--Huya--Sun angle) was $\bar{\alpha}$ = 1\fdg8. The periodogram in Fig. \ref{huy2006}, middle panel, shows several minima. Our 
      best fit for the rotational period is $P$ = 4.45$\pm$0.07 h (or a frequency of 0.225$\pm$0.003 rotations per hour). This is close to 
      one of the aliases in Thirouin et al. (2014), 4.31 h. The probability that there is no period with value $P$ is 32.5$\pm$2.1\% and 
      that of the observations containing a period that is different from $P$ is 1.0$\pm$0.4\%. The light-curve of Huya in Fig. 
      \ref{huy2006}, bottom panel, represents the detrended data from the top panel phased with the best-fit period. Our sparse curve looks 
      similar to that in fig. 11 of Thirouin et al. (2014), its amplitude is $\sim$0.1 mag. For Huya, we found the following values of the 
      colours: $B-V=1.00\pm0.06$, $V-R=0.58\pm0.09$, and $R-I=0.64\pm0.20$. These are compatible with those in e.g. Hainaut \& Delsanti 
      (2002).
%
%
%
%
%
%
     \begin{table}
        \fontsize{8}{11pt}\selectfont
        \tabcolsep 0.10truecm
        \caption{Photometry of 38628 Huya (2000 EB$_{173}$); 2006 June. Notation as in Table \ref{photometrychironjune2006}.}
        \centering
        \begin{tabular}{lccccc}
           \hline
           \hline
            \multicolumn{1}{c}{Julian date}      &
            \multicolumn{1}{c}{Filter}           &
            \multicolumn{1}{c}{A.M.}             &
            \multicolumn{1}{c}{Exp. (s)}         &
            \multicolumn{1}{c}{Mag}              &
            \multicolumn{1}{c}{$\alpha$ (\degr)} \\
           \hline
            2453914.494178 & $R$ & 1.15 & 400 & 19.209$\pm$0.029 & 1.82 \\
            2453914.506551 & $R$ & 1.14 & 400 & 19.168$\pm$0.034 & 1.82 \\
            2453914.519028 & $R$ & 1.13 & 400 & 19.167$\pm$0.032 & 1.82 \\
            2453914.552604 & $R$ & 1.15 & 400 & 19.199$\pm$0.038 & 1.82 \\
            2453914.558831 & $R$ & 1.15 & 400 & 19.207$\pm$0.035 & 1.82 \\
            2453914.592060 & $R$ & 1.24 & 400 & 19.173$\pm$0.040 & 1.82 \\
            2453914.598079 & $R$ & 1.26 & 400 & 19.193$\pm$0.038 & 1.82 \\
            2453914.624873 & $R$ & 1.39 & 400 & 19.263$\pm$0.041 & 1.82 \\
            2453914.643403 & $R$ & 1.53 & 400 & 19.204$\pm$0.046 & 1.82 \\
            2453914.649433 & $R$ & 1.59 & 400 & 19.275$\pm$0.048 & 1.82 \\
            2453916.471181 & $R$ & 1.18 & 400 & 19.351$\pm$0.045 & 1.84 \\
            2453916.477211 & $R$ & 1.17 & 400 & 19.359$\pm$0.042 & 1.84 \\
            2453916.483241 & $R$ & 1.16 & 400 & 19.274$\pm$0.038 & 1.84 \\
            2453916.518438 & $R$ & 1.13 & 500 & 19.352$\pm$0.042 & 1.84 \\
            2453916.543808 & $R$ & 1.14 & 600 & 19.253$\pm$0.035 & 1.84 \\
            2453916.569086 & $R$ & 1.18 & 600 & 19.255$\pm$0.039 & 1.84 \\
            2453916.595590 & $R$ & 1.26 & 600 & 19.351$\pm$0.037 & 1.85 \\
            2453916.621944 & $R$ & 1.40 & 600 & 19.279$\pm$0.070 & 1.85 \\
            2453916.663704 & $R$ & 1.63 & 600 & 20.790$\pm$0.578 & 1.85 \\
           \hline
            2453914.500347 & $V$ & 1.14 & 400 & 19.784$\pm$0.023 & 1.82 \\
            2453914.631030 & $V$ & 1.43 & 400 & 19.791$\pm$0.029 & 1.82 \\
           \hline
            2453914.512812 & $I$ & 1.13 & 400 & 18.575$\pm$0.028 & 1.82 \\
            2453914.637211 & $I$ & 1.48 & 400 & 18.564$\pm$0.026 & 1.82 \\
           \hline
        \end{tabular}
        \label{huya}
     \end{table}
%
%

   \section{28978 Ixion (2001 KX$_{76}$)}
\begin{figure}
       \centering
        \includegraphics[width=\linewidth]{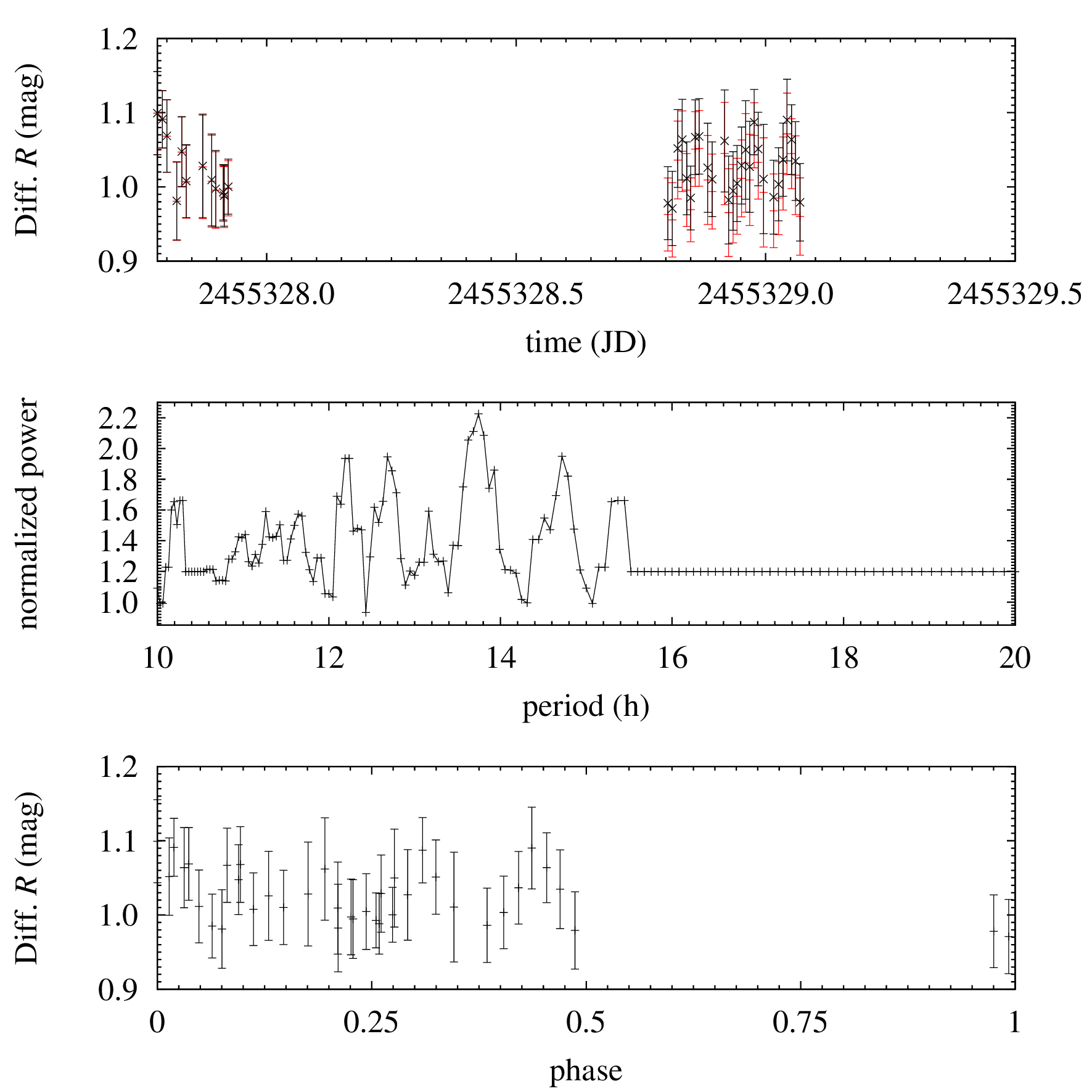}
        \caption{Same as Fig. \ref{chi2006} but for Ixion, data obtained in 2010 May; 4$\times$39 test frequencies. The lowest value of the 
                 normalized power corresponds to a rotation period of 12.4$\pm$0.3 h or a frequency of 0.080$\pm$0.002 cycles/h with a 
                 false alarm probability $<$ 2\%. The bottom panel shows the rotational light-curve of Ixion phased to a period of 12.4 h. 
                }
        \label{ixion2010}
     \end{figure}

      Ixion is also a Plutino, one of the largest known. Lellouch et al. (2013) give a value of the diameter of 617$\pm$20~km, with an 
      albedo of 14.1$\pm$1.1\%. A rotational period of 15.9 h has been derived by Rousselot \& Petit (2010). Ixion was observed in 2010 May 
      (see Tables \ref{ix10} and \ref{astrometryixion2010}) when the object was in Ophiuchus with the same equipment used for the previous 
      three objects. The average geocentric distance of Ixion during the observing run was $\bar{\Delta}$ = 40.48 AU, the average 
      heliocentric distance was $\bar{r}$=41.38~AU, and the average phase angle (Earth--Ixion--Sun angle) was $\bar{\alpha}$ = 0\fdg6. The 
      periodogram in Fig. \ref{ixion2010}, middle panel, shows several minima. Our best fit for the rotational period is $P$ = 12.4$\pm$0.3 
      h (or a frequency of 0.080$\pm$0.002 rotations per hour). The light-curve is rather flat and the error bars are nearly as large as the 
      purported photometric amplitude. The probability that there is no period with value $P$ is 1.2$\pm$0.5\% and that of the observations 
      containing a period that is different from $P$ is $<$0.01\%. The light-curve of Ixion in Fig. \ref{ixion2010}, bottom panel, 
      represents the detrended data from the top panel phased with the best-fit period. Unfortunately, the data sampling was rather 
      incomplete. No evidence of cometary activity was found in our CCD frames; this result is consistent with those from previous studies 
      (e.g. Lorin \& Rousselot 2007). Ixion was also observed in 2006 (see Table \ref{astrometryixion2006}), but no useful data were 
      acquired other than astrometry. The average geocentric distance of Ixion during this previous observing run was 
      $\bar{\Delta}$=41.34~AU, the average heliocentric distance was $\bar{r}$=42.25~AU, and the average phase angle was 
      $\bar{\alpha}$=0\fdg6.
%
%
%
%
%
%
     \begin{table}
        \fontsize{8}{11pt}\selectfont
        \tabcolsep 0.20truecm
        \caption{Photometry of 28978 Ixion (2001 KX$_{76}$); 2010 May. All the observations in the $R$ filter; differential magnitudes are 
                 used.}
        \centering
        \begin{tabular}{lc}
           \hline
           \hline
            \multicolumn{1}{c}{Julian date}            &
            \multicolumn{1}{c}{Mag}                    \\
           \hline
            2455327.7814 & 1.099$\pm$0.056 \\
            2455327.7913 & 1.091$\pm$0.039 \\
            2455327.7997 & 1.069$\pm$0.049 \\
            2455327.8198 & 0.981$\pm$0.053 \\
            2455327.8297 & 1.048$\pm$0.047 \\
            2455327.8387 & 1.008$\pm$0.049 \\
            2455327.8723 & 1.028$\pm$0.070 \\
            2455327.8897 & 1.009$\pm$0.062 \\
            2455327.8982 & 0.997$\pm$0.052 \\
            2455327.9128 & 0.993$\pm$0.037 \\
            2455327.9150 & 0.988$\pm$0.041 \\
            2455327.9234 & 1.000$\pm$0.037 \\
            2455328.8044 & 0.978$\pm$0.049 \\
            2455328.8132 & 0.971$\pm$0.050 \\
            2455328.8241 & 1.052$\pm$0.052 \\
            2455328.8329 & 1.064$\pm$0.054 \\
            2455328.8419 & 1.011$\pm$0.049 \\
            2455328.8504 & 0.985$\pm$0.043 \\
            2455328.8588 & 1.067$\pm$0.050 \\
            2455328.8672 & 1.068$\pm$0.051 \\
            2455328.8840 & 1.026$\pm$0.060 \\
            2455328.8925 & 1.010$\pm$0.050 \\
            2455328.9177 & 1.062$\pm$0.069 \\
            2455328.9261 & 0.982$\pm$0.059 \\
            2455328.9346 & 0.995$\pm$0.053 \\
            2455328.9431 & 1.005$\pm$0.051 \\
            2455328.9515 & 1.029$\pm$0.052 \\
            2455328.9599 & 1.050$\pm$0.066 \\
            2455328.9683 & 1.027$\pm$0.061 \\
            2455328.9767 & 1.087$\pm$0.044 \\
            2455328.9852 & 1.051$\pm$0.050 \\
            2455328.9959 & 1.011$\pm$0.074 \\
            2455329.0164 & 0.986$\pm$0.050 \\
            2455329.0263 & 1.003$\pm$0.049 \\
            2455329.0347 & 1.037$\pm$0.049 \\
            2455329.0431 & 1.090$\pm$0.055 \\
            2455329.0516 & 1.064$\pm$0.047 \\
            2455329.0600 & 1.035$\pm$0.053 \\
            2455329.0685 & 0.979$\pm$0.052 \\
           \hline
        \end{tabular}
        \label{ix10}
     \end{table}
%
%

   \section{90482 Orcus (2004 DW)}
\begin{figure}
       \centering
        \includegraphics[width=\linewidth]{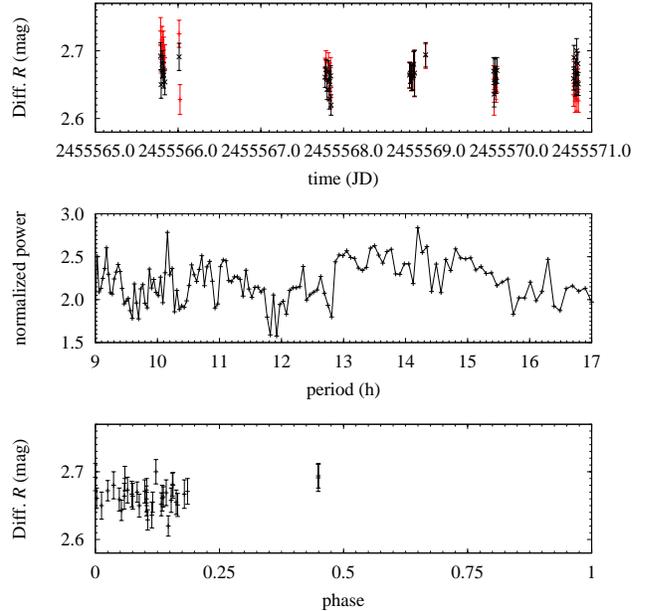}
        \caption{Same as Fig. \ref{chi2006} but for Orcus, data obtained in 2011 January; 4$\times$36 test frequencies. The lowest value of 
                 the normalized power corresponds to a rotation period of 11.9$\pm$0.5 h or a frequency of 0.084$\pm$0.004 cycles/h with 
                 a false alarm probability $>$ 90\%. The bottom panel shows the rotational light-curve of Orcus phased to a period of 11.9 
                 h.}
        \label{orc2011}
     \end{figure}

      Also a Plutino, Orcus is the largest of the objects studied in this paper. Its diameter amounts to 917$\pm$25~km with an albedo of 
      23$\pm$2\% (Fornasier et al. 2013). Rabinowitz et al. (2007) give a value of 13.188 h for the rotational period of this object. 
      Fornasier et al. (2013) suggest 10.47 h based on Thirouin et al. (2010) that gives an amplitude of 0.04$\pm$0.01~mag; a similar
      estimate is also given in Ortiz et al. (2006). Orcus has a known companion, Vanth, whose mass is comparable to that of the primary 
      (Brown et al. 2010); its diameter is estimated to be 276$\pm$17~km (Fornasier et al. 2013). This may induce tidally locked rotation in 
      the pair and Ortiz et al. (2011) have found possible evidence of this in the form of a photometric variability with a period of 
      9.7$\pm$0.3 days. Orcus was observed in 2010 May (see Tables \ref{orc10} and \ref{astrometryorcusmay2010}) and again in 2011 January 
      (see Tables \ref{orc11} and \ref{astrometryorcusjanuary2011}) when it was in Sextans with the same equipment used for the previous 
      four objects. The average geocentric distance of Orcus during the first observing run was $\bar{\Delta}$ = 47.73 AU, the average 
      heliocentric distance was $\bar{r}$=47.90~AU, and the average phase angle (Earth--Orcus--Sun angle) was $\bar{\alpha}$ = 1\fdg2; the 
      respective values for the second observing run were 47.29 AU, 47.92~AU and 0\fdg9. The periodogram in Fig. \ref{orc2011}, middle 
      panel, results from the analysis of the second run and shows several minima. Our best fit for the rotational period is $P$ = 
      11.9$\pm$0.5 h (or a frequency of 0.084$\pm$0.004 rotations per hour). Unfortunately, the light-curve is rather flat and incomplete; 
      it resembles those in figs. 18 and 19 in Sheppard (2007) that show an amplitude $<0.03$~mag. The probability that there is no period 
      with value $P$ is $>$ 90\%. The light-curve of Orcus in Fig. \ref{orc2011}, bottom panel, represents the data from the top panel 
      phased with the best-fit period and exhibits a photometric amplitude of $\sim$0.05 which is similar in value to the error bars and 
      also to the values cited in the literature. Small amplitude light-curves are characteristic of spherical objects with featureless
      surfaces and/or those observed under a nearly pole-on viewing geometry.
%
%
%
%
%
%
     \begin{table}
        \fontsize{8}{11pt}\selectfont
        \tabcolsep 0.20truecm
        \caption{Photometry of 90482 Orcus (2004 DW); 2010 May. All the observations in the $R$ filter; differential magnitudes are used.}
        \centering
        \begin{tabular}{lccc}
           \hline
           \hline
            \multicolumn{1}{c}{Julian date}            &
            \multicolumn{1}{c}{A.M.}                   &
            \multicolumn{1}{c}{Mag}                    &
            \multicolumn{1}{c}{$\alpha$(\degr)}        \\
           \hline
            2455327.4600 & 1.09 & 0.484$\pm$0.017 & 1.19 \\
            2455327.4684 & 1.09 & 0.506$\pm$0.016 & 1.19 \\
            2455327.4769 & 1.09 & 0.513$\pm$0.017 & 1.19 \\
            2455327.4854 & 1.10 & 0.536$\pm$0.014 & 1.19 \\
            2455327.4938 & 1.10 & 0.511$\pm$0.015 & 1.19 \\
            2455327.5028 & 1.12 & 0.524$\pm$0.017 & 1.19 \\
            2455327.5112 & 1.13 & 0.525$\pm$0.017 & 1.19 \\
            2455327.5196 & 1.15 & 0.496$\pm$0.013 & 1.19 \\
            2455327.5280 & 1.17 & 0.495$\pm$0.012 & 1.19 \\
            2455327.5364 & 1.20 & 0.521$\pm$0.013 & 1.19 \\
            2455327.5468 & 1.23 & 0.535$\pm$0.015 & 1.19 \\
            2455327.5552 & 1.27 & 0.527$\pm$0.013 & 1.19 \\
            2455327.5637 & 1.32 & 0.529$\pm$0.013 & 1.19 \\
            2455327.5721 & 1.37 & 0.515$\pm$0.011 & 1.19 \\
            2455327.5862 & 1.47 & 0.527$\pm$0.014 & 1.19 \\
            2455327.5947 & 1.54 & 0.547$\pm$0.015 & 1.19 \\
            2455327.6031 & 1.62 & 0.545$\pm$0.015 & 1.19 \\
           \hline
        \end{tabular}
        \label{orc10}
     \end{table}
%
%
%
%
     \begin{table}
        \fontsize{8}{11pt}\selectfont
        \tabcolsep 0.20truecm
        \caption{Photometry of 90482 Orcus (2004 DW); 2011 January. All the observations in the $R$ filter; differential magnitudes are 
                 used.}
        \centering
        \begin{tabular}{lccc}
           \hline
           \hline
            \multicolumn{1}{c}{Julian date}            &
            \multicolumn{1}{c}{A.M.}                   &
            \multicolumn{1}{c}{Mag}                    &
            \multicolumn{1}{c}{$\alpha$(\degr)}        \\
           \hline
            2455565.7892 & 1.10 & 2.729$\pm$0.020 & 0.93 \\
            2455565.7954 & 1.09 & 2.687$\pm$0.020 & 0.93 \\
            2455565.8074 & 1.08 & 2.716$\pm$0.020 & 0.93 \\
            2455565.8127 & 1.08 & 2.552$\pm$0.025 & 0.93 \\
            2455565.8214 & 1.08 & 2.709$\pm$0.020 & 0.93 \\
            2455565.8266 & 1.08 & 2.701$\pm$0.019 & 0.92 \\
            2455565.8404 & 1.09 & 2.690$\pm$0.019 & 0.92 \\
            2455566.0123 & 1.09 & 2.725$\pm$0.020 & 0.92 \\
            2455566.0234 & 1.10 & 2.628$\pm$0.022 & 0.92 \\
            2455566.0284 & 1.11 & 2.804$\pm$0.029 & 0.92 \\
            2455567.7775 & 1.11 & 2.674$\pm$0.015 & 0.90 \\
            2455567.7886 & 1.09 & 2.685$\pm$0.015 & 0.90 \\
            2455567.8023 & 1.08 & 2.655$\pm$0.015 & 0.90 \\
            2455567.8180 & 1.08 & 2.682$\pm$0.015 & 0.90 \\
            2455567.8286 & 1.08 & 2.641$\pm$0.015 & 0.90 \\
            2455567.8336 & 1.08 & 2.667$\pm$0.015 & 0.90 \\
            2455567.8442 & 1.10 & 2.675$\pm$0.015 & 0.90 \\
            2455567.8492 & 1.10 & 2.631$\pm$0.015 & 0.90 \\
            2455568.7987 & 1.08 & 2.664$\pm$0.019 & 0.89 \\
            2455568.8062 & 1.08 & 2.667$\pm$0.015 & 0.89 \\
            2455568.8209 & 1.08 & 2.660$\pm$0.019 & 0.89 \\
            2455568.9929 & 1.08 & 2.692$\pm$0.018 & 0.89 \\
            2455568.8367 & 1.09 & 2.661$\pm$0.019 & 0.89 \\
            2455568.8474 & 1.10 & 2.680$\pm$0.019 & 0.89 \\
            2455568.8588 & 1.12 & 2.666$\pm$0.034 & 0.89 \\
            2455569.8146 & 1.08 & 2.658$\pm$0.020 & 0.88 \\
            2455569.8198 & 1.08 & 2.624$\pm$0.019 & 0.88 \\
            2455569.8341 & 1.09 & 2.657$\pm$0.019 & 0.88 \\
            2455569.8391 & 1.10 & 2.646$\pm$0.018 & 0.88 \\
            2455569.8442 & 1.10 & 2.642$\pm$0.018 & 0.88 \\
            2455569.8555 & 1.12 & 2.658$\pm$0.019 & 0.88 \\
            2455570.7807 & 1.09 & 2.635$\pm$0.017 & 0.87 \\
            2455570.7863 & 1.09 & 2.666$\pm$0.018 & 0.87 \\
            2455570.8007 & 1.08 & 2.626$\pm$0.017 & 0.87 \\
            2455570.8061 & 1.08 & 2.647$\pm$0.017 & 0.87 \\
            2455570.8174 & 1.08 & 2.676$\pm$0.018 & 0.87 \\
            2455570.8227 & 1.08 & 2.628$\pm$0.017 & 0.87 \\
            2455570.8339 & 1.09 & 2.656$\pm$0.017 & 0.87 \\
            2455570.8391 & 1.10 & 2.626$\pm$0.017 & 0.87 \\
           \hline
        \end{tabular}
        \label{orc11}
     \end{table}
%
%

   \section{Discussion and conclusions}
      We have collected and analysed $R$-band photometric data for two Centaurs and three TNOs. In principle, our analysis confirms the 
      published values of the rotational periods of Chiron, Chariklo, and Huya; the photometric amplitudes found are, in general, consistent 
      with those quoted in the literature. These also exhibit notable dispersions, in particular those of Chiron. This may hint at changing 
      surface features or, perhaps, chaotic rotation. Both Ixion and Orcus show behaviour compatible with no variability within the 
      photometric uncertainties. Assuming that the data are reliable, lack of brightness variation may have its origin in slow spin, being 
      viewed nearly pole-on, and/or round shape. In general, our rotational period results could be uncertain by a few tens of percent as 
      they are based on less than full coverage of the light-curve. As for the overall rotational properties of Centaurs and TNOs, the
      extensive analysis in Thirouin et al. (2014) shows that single TNOs tend to spin faster than binaries. On the other hand, resonant
      TNOs (Plutinos in particular) are less prone to suffer planetary close encounters, such dynamical events may alter the rotational
      properties of Centaurs that are more likely to experience tidal interactions with the Jovian planets. In this context, the current
      values of the rotational periods of Chiron and Chariklo may not be primordial.

\begin{figure}
       \centering
        \includegraphics[width=\linewidth]{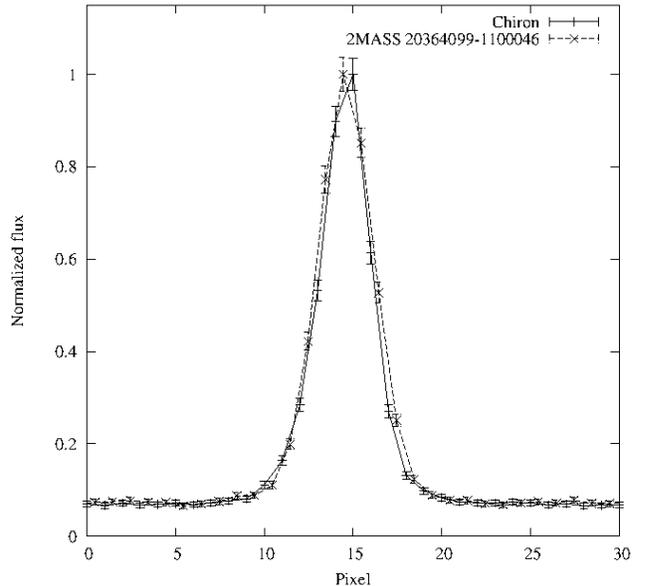}
        \caption{The brightness profile of 2060 Chiron and that of a scaled background star (2MASS 20364099-1100046, $B$ = 18.27,
                 $V$ = 17.17, $R$ = 17.23) illustrate the stellar appearance of the Centaur.
                }
        \label{PSFChiron}
     \end{figure}

      Visual inspection and measurements of the FWHM of the objects studied here and neighbouring star images did not reveal the presence of 
      a coma around them. As an example, Fig. \ref{PSFChiron} compares the brightness profile of Chiron (first run) with that of a scaled 
      background star. Their profiles are indistinguishable in all directions. Since the radial profile of Chiron is basically identical to 
      that of the comparison stars, we conclude that a coma around Chiron was not present or it was well beyond our detection limit (order 
      of 27.18 mag/arcsec$^2$) ---if present at all. The absence of a detectable coma is compatible with the results in Fornasier et al. 
      (2013).  In order to constrain the possible presence of a coma, we use the relation given by Jewitt \& Danielson (1984)
      \begin{equation}
         \Sigma(\phi) = m(\phi) + 2.5 \log (2 \pi \phi^{2})\,,
      \end{equation}
      where $m(\phi)$ is the total magnitude of the coma inside a circle of radius $\phi$ in arcsec and $\Sigma(\phi)$ is the surface 
      brightness at projected radius $\phi$. The upper limit to the surface brightness of a hypothetical coma around the object at $\phi = 
      2\farcs175$ (almost double the seeing) can be set to be 27.18 mag/arcsec$^2$ as $m(2\farcs175) >$ 20 mag in $R$. This is 3.49 mag 
      fainter (factor of 25) than the limiting value for a single frame as 25 of them were coadded; for a SNR of 10 the limiting magnitude 
      is 20 mag. Consistently, presence of candidate satellites or comoving wide companions brighter than 23.5 mag in $R$ can also be 
      discarded. Similar results have been obtained for the other objects. Modelling the dust production rate for Chiron or Chariklo is 
      outside the scope of this work. The orbital solutions derived from the acquired astrometry (see Appendix A) are compatible with those 
      already available from the JPL Small-Body Database, the MPC data server, or the AstDyS information service. 
%
%
%
%

   \acknowledgments
      We thank the anonymous referee for her/his constructive and helpful reports. M. G. and G. C. express their gratitude to S. Ortolani 
      and S. Marchi for many useful discussions and advice. We thank R. Mateluna and L. J\'{\i}lkov\'a for their help with the observations. 
      In preparation of this paper, we made use of the NASA Astrophysics Data System, the ASTRO-PH e-print server, the JPL Small-Body 
      Database, the MPC data server, and the AstDyS information service.

  \newpage
  \appendix
  \section{Astrometry of 2060 Chiron (1977 UB), 10199 Chariklo (1997 CU$_{26}$), 38628 Huya (2000 EB$_{173}$), 28978 Ixion
          (2001 KX$_{76}$), and 90482 Orcus (2004 DW)}
%
%
     \begin{table}
        \fontsize{8}{11pt}\selectfont
        \tabcolsep 0.20truecm
        \caption{Observations of 2060 Chiron (1977 UB) over interval 2006 June 26.298391 -- 28.313276. During this observing run, Chiron was 
                 in Capricornus. See also Table \ref{photometrychironjune2006}.}
        \centering
        \begin{tabular}{cccc}
           \hline
           \hline
            Date             & RA(J2000)                          & Dec(J2000)              & Filter \\
            (UT)             & ($^{\rm h}$:$^{\rm m}$:$^{\rm s}$) & (\degr:\arcmin:\arcsec) &        \\
           \hline
            2006 6 26.298391 & 20:36:46.81                        & $-$11:00:29.8           & $B$    \\
            2006 6 26.305822 & 20:36:46.72                        & $-$11:00:30.1           & $R$    \\
            2006 6 26.311991 & 20:36:46.67                        & $-$11:00:30.3           & $V$    \\
            2006 6 26.318206 & 20:36:46.61                        & $-$11:00:30.4           & $I$    \\
            2006 6 26.324410 & 20:36:46.54                        & $-$11:00:30.5           & $R$    \\
            2006 6 26.357975 & 20:36:46.18                        & $-$11:00:31.0           & $R$    \\
            2006 6 26.364005 & 20:36:46.12                        & $-$11:00:31.2           & $R$    \\
            2006 6 26.370023 & 20:36:46.04                        & $-$11:00:31.4           & $R$    \\
            2006 6 26.376053 & 20:36:45.99                        & $-$11:00:31.6           & $R$    \\
            2006 6 26.382084 & 20:36:45.93                        & $-$11:00:31.8           & $R$    \\
            2006 6 26.389491 & 20:36:45.86                        & $-$11:00:31.8           & $B$    \\
            2006 6 26.397037 & 20:36:45.78                        & $-$11:00:31.9           & $R$    \\
            2006 6 26.403067 & 20:36:45.69                        & $-$11:00:31.9           & $R$    \\
            2006 6 26.410428 & 20:36:45.64                        & $-$11:00:32.4           & $B$    \\
            2006 6 28.189248 & 20:36:26.98                        & $-$11:01:13.5           & $R$    \\
            2006 6 28.195278 & 20:36:26.92                        & $-$11:01:13.7           & $R$    \\
            2006 6 28.201296 & 20:36:26.84                        & $-$11:01:13.8           & $R$    \\
            2006 6 28.224491 & 20:36:26.57                        & $-$11:01:14.5           & $R$    \\
            2006 6 28.230521 & 20:36:26.51                        & $-$11:01:14.7           & $R$    \\
            2006 6 28.236540 & 20:36:26.43                        & $-$11:01:15.0           & $R$    \\
            2006 6 28.258970 & 20:36:26.20                        & $-$11:01:15.3           & $R$    \\
            2006 6 28.265000 & 20:36:26.14                        & $-$11:01:15.5           & $R$    \\
            2006 6 28.271030 & 20:36:26.07                        & $-$11:01:15.8           & $R$    \\  
            2006 6 28.277049 & 20:36:26.00                        & $-$11:01:16.0           & $R$    \\
            2006 6 28.283079 & 20:36:25.94                        & $-$11:01:16.2           & $R$    \\
            2006 6 28.289167 & 20:36:25.87                        & $-$11:01:16.2           & $R$    \\
            2006 6 28.294503 & 20:36:25.81                        & $-$11:01:16.4           & $R$    \\
            2006 6 28.301227 & 20:36:25.74                        & $-$11:01:16.5           & $R$    \\
            2006 6 28.307246 & 20:36:25.70                        & $-$11:01:16.6           & $R$    \\
            2006 6 28.313276 & 20:36:25.59                        & $-$11:01:16.8           & $R$    \\
           \hline
        \end{tabular}
        \label{astrometrychironjune2006}
     \end{table}
%
%
%
%
     \begin{table}
        \fontsize{8}{11pt}\selectfont
        \tabcolsep 0.20truecm
        \caption{Observations of 2060 Chiron (1977 UB) over interval 2011 July 30.217766 -- August 06.393033. During this observing run, 
                 Chiron was in Aquarius. All the observations in the $R$ filter. See also Table \ref{photometrychiron2011}.}
        \centering
        \resizebox{0.40\linewidth}{0.68\linewidth}{
        \begin{tabular}{ccc}
           \hline
           \hline
            Date             & RA(J2000)                          & Dec(J2000)              \\
            (UT)             & ($^{\rm h}$:$^{\rm m}$:$^{\rm s}$) & (\degr:\arcmin:\arcsec) \\
           \hline
            2011 7 30.217766 & 22:15:08.78                        & $-$04:16:07.4           \\
            2011 7 30.226655 & 22:15:08.68                        & $-$04:16:07.8           \\
            2011 7 30.238704 & 22:15:08.57                        & $-$04:16:08.2           \\
            2011 7 30.250452 & 22:15:08.43                        & $-$04:16:09.0           \\
            2011 7 30.261412 & 22:15:08.35                        & $-$04:16:09.4           \\
            2011 7 30.272199 & 22:15:08.24                        & $-$04:16:09.9           \\
            2011 7 30.283102 & 22:15:08.12                        & $-$04:16:10.6           \\
            2011 7 30.293854 & 22:15:08.02                        & $-$04:16:11.2           \\
            2011 7 30.304618 & 22:15:07.91                        & $-$04:16:11.6           \\
            2011 7 30.315347 & 22:15:07.82                        & $-$04:16:12.4           \\
            2011 7 30.329169 & 22:15:07.68                        & $-$04:16:12.8           \\
            2011 7 30.340347 & 22:15:07.58                        & $-$04:16:13.2           \\
            2011 7 30.350047 & 22:15:07.47                        & $-$04:16:13.8           \\
            2011 7 30.358658 & 22:15:07.42                        & $-$04:16:14.1           \\
            2011 7 30.367060 & 22:15:07.30                        & $-$04:16:14.7           \\
            2011 7 30.375475 & 22:15:07.20                        & $-$04:16:15.0           \\
            2011 7 30.383889 & 22:15:07.15                        & $-$04:16:15.5           \\
            2011 7 30.392292 & 22:15:07.04                        & $-$04:16:15.8           \\
            2011 7 30.400706 & 22:15:06.96                        & $-$04:16:16.2           \\
            2011 7 30.409144 & 22:15:06.87                        & $-$04:16:16.5           \\
            2011 7 30.417558 & 22:15:06.80                        & $-$04:16:17.1           \\
            2011 8 02.194271 & 22:14:39.86                        & $-$04:18:31.1           \\
            2011 8 02.202766 & 22:14:39.77                        & $-$04:18:31.7           \\
            2011 8 02.212813 & 22:14:39.67                        & $-$04:18:32.3           \\
            2011 8 02.223380 & 22:14:39.57                        & $-$04:18:32.8           \\
            2011 8 02.246100 & 22:14:39.35                        & $-$04:18:33.8           \\
            2011 8 02.258449 & 22:14:39.21                        & $-$04:18:34.4           \\
            2011 8 02.270695 & 22:14:39.08                        & $-$04:18:35.4           \\
            2011 8 02.292616 & 22:14:38.86                        & $-$04:18:36.5           \\
            2011 8 02.305799 & 22:14:38.72                        & $-$04:18:37.0           \\
            2011 8 02.316470 & 22:14:38.63                        & $-$04:18:37.6           \\
            2011 8 02.360116 & 22:14:38.18                        & $-$04:18:39.6           \\
            2011 8 02.371898 & 22:14:38.06                        & $-$04:18:40.4           \\
            2011 8 02.380336 & 22:14:37.99                        & $-$04:18:40.8           \\
            2011 8 02.388762 & 22:14:37.88                        & $-$04:18:41.3           \\
            2011 8 02.397188 & 22:14:37.80                        & $-$04:18:41.8           \\
            2011 8 02.405614 & 22:14:37.70                        & $-$04:18:42.2           \\
            2011 8 02.414040 & 22:14:37.62                        & $-$04:18:42.7           \\
            2011 8 06.179375 & 22:13:59.81                        & $-$04:21:57.2           \\
            2011 8 06.187778 & 22:13:59.71                        & $-$04:21:57.7           \\
            2011 8 06.197523 & 22:13:59.61                        & $-$04:21:58.3           \\
            2011 8 06.205926 & 22:13:59.55                        & $-$04:21:58.8           \\
            2011 8 06.214317 & 22:13:59.43                        & $-$04:21:59.2           \\
            2011 8 06.222847 & 22:13:59.34                        & $-$04:21:59.7           \\
            2011 8 06.231239 & 22:13:59.22                        & $-$04:21:59.9           \\
            2011 8 06.239641 & 22:13:59.13                        & $-$04:22:00.4           \\
            2011 8 06.265116 & 22:13:58.96                        & $-$04:22:01.7           \\
            2011 8 06.273600 & 22:13:58.84                        & $-$04:22:02.4           \\
            2011 8 06.281991 & 22:13:58.74                        & $-$04:22:02.8           \\
            2011 8 06.290394 & 22:13:58.64                        & $-$04:22:03.2           \\
            2011 8 06.314792 & 22:13:58.39                        & $-$04:22:04.3           \\
            2011 8 06.325417 & 22:13:58.27                        & $-$04:22:04.8           \\
            2011 8 06.333854 & 22:13:58.19                        & $-$04:22:05.2           \\
            2011 8 06.342246 & 22:13:58.12                        & $-$04:22:05.8           \\
            2011 8 06.350695 & 22:13:58.03                        & $-$04:22:06.2           \\
            2011 8 06.359097 & 22:13:57.94                        & $-$04:22:06.8           \\
            2011 8 06.367489 & 22:13:57.85                        & $-$04:22:07.3           \\
            2011 8 06.376146 & 22:13:57.77                        & $-$04:22:07.7           \\
            2011 8 06.384560 & 22:13:57.68                        & $-$04:22:08.1           \\
            2011 8 06.393033 & 22:13:57.59                        & $-$04:22:08.7           \\
            2011 8 06.401435 & 22:13:57.50                        & $-$04:22:09.0           \\
           \hline
        \end{tabular}
        }
        \label{astrometrychiron2011}
     \end{table}
%
%
%
%
     \begin{table}
        \fontsize{8}{11pt}\selectfont
        \tabcolsep 0.20truecm
        \caption{Observations of 2060 Chiron (1977 UB) over interval 2006 May 19.383990 -- 22.438617. During this observing run, Chiron was 
                 in Aquarius.}
        \centering
        \begin{tabular}{cccc}
           \hline
           \hline
            Date             & RA(J2000)                          & Dec(J2000)              & Filter \\
            (UT)             & ($^{\rm h}$:$^{\rm m}$:$^{\rm s}$) & (\degr:\arcmin:\arcsec) &        \\
           \hline
            2006 5 19.383990 & 20:40:46.42                        & $-$11:00:36.5           & $R$    \\
            2006 5 19.389247 & 20:40:46.41                        & $-$11:00:36.2           & $V$    \\
            2006 5 19.394501 & 20:40:46.39                        & $-$11:00:36.2           & $R$    \\
            2006 5 19.407375 & 20:40:46.39                        & $-$11:00:35.9           & $R$    \\
            2006 5 19.412628 & 20:40:46.37                        & $-$11:00:35.9           & $I$    \\
            2006 5 19.417883 & 20:40:46.34                        & $-$11:00:35.6           & $R$    \\
            2006 5 19.423152 & 20:40:46.33                        & $-$11:00:35.5           & $V$    \\
            2006 5 19.428409 & 20:40:46.32                        & $-$11:00:35.4           & $R$    \\
            2006 5 19.441269 & 20:40:46.30                        & $-$11:00:35.1           & $R$    \\
            2006 5 19.446528 & 20:40:46.28                        & $-$11:00:35.0           & $I$    \\
            2006 5 22.426946 & 20:40:40.18                        & $-$10:59:32.1           & $R$    \\
            2006 5 22.432782 & 20:40:40.18                        & $-$10:59:31.9           & $V$    \\
            2006 5 22.438617 & 20:40:40.16                        & $-$10:59:31.8           & $R$    \\
           \hline
        \end{tabular}
        \label{astrometrychironmay2006}
     \end{table}
%
%
%
%
     \begin{table}
        \fontsize{8}{11pt}\selectfont
        \tabcolsep 0.20truecm
        \caption{Observations of Chariklo over interval 2006 June 27.950405 -- 30.113403. See also Table \ref{run06}. During this observing 
                 run, Chariklo was in Hydra.}
        \centering
        \begin{tabular}{cccc}
           \hline
           \hline
            Date             & RA(J2000)                          & Dec(J2000)              & Filter \\
            (UT)             & ($^{\rm h}$:$^{\rm m}$:$^{\rm s}$) & (\degr:\arcmin:\arcsec) &        \\
           \hline
            2006 6 27.950405 & 12:24:30.52                        & $-$28:28:12.9           & $R$    \\
            2006 6 27.956100 & 12:24:30.54                        & $-$28:28:12.5           & $V$    \\
            2006 6 27.962315 & 12:24:30.57                        & $-$28:28:12.2           & $R$    \\
            2006 6 27.968484 & 12:24:30.59                        & $-$28:28:11.6           & $V$    \\
            2006 6 27.974676 & 12:24:30.62                        & $-$28:28:11.3           & $R$    \\
            2006 6 27.980833 & 12:24:30.65                        & $-$28:28:10.7           & $I$    \\
            2006 6 27.987072 & 12:24:30.68                        & $-$28:28:10.3           & $R$    \\
            2006 6 28.039294 & 12:24:30.93                        & $-$28:28:06.2           & $R$    \\
            2006 6 28.045313 & 12:24:30.96                        & $-$28:28:05.8           & $R$    \\
            2006 6 28.078819 & 12:24:31.12                        & $-$28:28:03.2           & $R$    \\
            2006 6 28.084850 & 12:24:31.14                        & $-$28:28:02.7           & $R$    \\
            2006 6 28.117986 & 12:24:31.31                        & $-$28:28:00.1           & $R$    \\
            2006 6 28.973970 & 12:24:35.84                        & $-$28:26:55.3           & $R$    \\
            2006 6 28.989641 & 12:24:35.93                        & $-$28:26:53.9           & $R$    \\
            2006 6 29.953565 & 12:24:41.27                        & $-$28:25:42.5           & $R$    \\
            2006 6 29.958426 & 12:24:41.30                        & $-$28:25:42.1           & $R$    \\
            2006 6 29.963299 & 12:24:41.32                        & $-$28:25:41.8           & $R$    \\
            2006 6 30.010891 & 12:24:41.58                        & $-$28:25:38.3           & $R$    \\
            2006 6 30.035081 & 12:24:41.70                        & $-$28:25:36.7           & $R$    \\
            2006 6 30.060914 & 12:24:41.85                        & $-$28:25:34.8           & $R$    \\
            2006 6 30.086921 & 12:24:41.99                        & $-$28:25:32.8           & $R$    \\
            2006 6 30.113403 & 12:24:42.15                        & $-$28:25:30.7           & $R$    \\
           \hline
        \end{tabular}
        \label{astrometry2006}
     \end{table}
%
%
%
%
     \begin{table}
        \fontsize{8}{11pt}\selectfont
        \tabcolsep 0.20truecm
        \caption{Observations of Chariklo over interval 2011 July 30.043623 -- August 06.162685. During this observing run, Chariklo was in 
                 Lupus. All the observations in the $R$ filter. See also Table \ref{run11}.}
        \centering
        \resizebox{0.40\linewidth}{0.65\linewidth}{
        \begin{tabular}{ccc}
           \hline
           \hline
            Date             & RA(J2000)                          & Dec(J2000)              \\
            (UT)             & ($^{\rm h}$:$^{\rm m}$:$^{\rm s}$) & (\degr:\arcmin:\arcsec) \\
           \hline
            2011 7 30.043623 & 15:29:12.14                        & $-$40:01:58.6           \\
            2011 7 30.052083 & 15:29:12.13                        & $-$40:01:57.7           \\
            2011 7 30.060567 & 15:29:12.13                        & $-$40:01:56.9           \\
            2011 7 30.074074 & 15:29:12.09                        & $-$40:01:55.4           \\
            2011 7 30.084757 & 15:29:12.09                        & $-$40:01:53.9           \\
            2011 7 30.100174 & 15:29:12.08                        & $-$40:01:52.8           \\
            2011 7 30.111632 & 15:29:12.07                        & $-$40:01:51.6           \\
            2011 7 30.122697 & 15:29:12.06                        & $-$40:01:50.6           \\
            2011 7 30.134063 & 15:29:12.06                        & $-$40:01:49.5           \\
            2011 7 30.145208 & 15:29:12.05	                  & $-$40:01:48.2           \\
            2011 7 30.156111 & 15:29:12.05                        & $-$40:01:47.1           \\
            2011 8 01.994016 & 15:29:12.71                        & $-$39:57:04.5           \\
            2011 8 02.002535 & 15:29:12.72                        & $-$39:57:03.7           \\
            2011 8 02.010972 & 15:29:12.72                        & $-$39:57:02.9           \\
            2011 8 02.019398 & 15:29:12.73                        & $-$39:57:02.0           \\
            2011 8 02.027824 & 15:29:12.73                        & $-$39:57:01.2           \\
            2011 8 02.036250 & 15:29:12.73                        & $-$39:57:00.4           \\
            2011 8 02.044699 & 15:29:12.73                        & $-$39:56:59.5           \\
            2011 8 02.053137 & 15:29:12.74                        & $-$39:56:58.8           \\
            2011 8 02.061563 & 15:29:12.74                        & $-$39:56:57.9           \\
            2011 8 02.069988 & 15:29:12.74                        & $-$39:56:56.9           \\
            2011 8 02.078426 & 15:29:12.75                        & $-$39:56:56.2           \\
            2011 8 02.086863 & 15:29:12.75                        & $-$39:56:55.3           \\
            2011 8 02.095289 & 15:29:12.76                        & $-$39:56:54.5           \\
            2011 8 02.103727 & 15:29:12.77                        & $-$39:56:53.7           \\
            2011 8 02.116771 & 15:29:12.77                        & $-$39:56:52.4           \\
            2011 8 02.126400 & 15:29:12.77                        & $-$39:56:51.5           \\
            2011 8 02.135914 & 15:29:12.78                        & $-$39:56:50.5           \\
            2011 8 02.144352 & 15:29:12.79                        & $-$39:56:49.6           \\
            2011 8 02.153252 & 15:29:12.79                        & $-$39:56:48.7           \\
            2011 8 02.161690 & 15:29:12.79                        & $-$39:56:47.8           \\
            2011 8 02.170139 & 15:29:12.80                        & $-$39:56:47.0           \\
            2011 8 02.178553 & 15:29:12.81                        & $-$39:56:46.2           \\
            2011 8 05.009792 & 15:29:16.44                        & $-$39:52:10.7           \\
            2011 8 05.018194 & 15:29:16.45                        & $-$39:52:09.9           \\
            2011 8 05.026620 & 15:29:16.47                        & $-$39:52:09.1           \\
            2011 8 05.036678 & 15:29:16.49                        & $-$39:52:08.3           \\
            2011 8 05.045590 & 15:29:16.50                        & $-$39:52:07.3           \\
            2011 8 05.073171 & 15:29:16.54                        & $-$39:52:04.5           \\
            2011 8 05.081586 & 15:29:16.55                        & $-$39:52:03.7           \\
            2011 8 05.119340 & 15:29:16.60                        & $-$39:52:00.0           \\
            2011 8 05.127755 & 15:29:16.62                        & $-$39:51:59.2           \\
            2011 8 05.136169 & 15:29:16.63                        & $-$39:51:58.4           \\
            2011 8 05.144676 & 15:29:16.65                        & $-$39:51:57.8           \\
            2011 8 05.153090 & 15:29:16.66                        & $-$39:51:57.2           \\
            2011 8 06.034132 & 15:29:18.44                        & $-$39:50:32.6           \\
            2011 8 06.043657 & 15:29:18.46                        & $-$39:50:31.7           \\
            2011 8 06.053125 & 15:29:18.47                        & $-$39:50:30.8           \\
            2011 8 06.061586 & 15:29:18.49                        & $-$39:50:29.9           \\
            2011 8 06.069988 & 15:29:18.51                        & $-$39:50:29.2           \\
            2011 8 06.078380 & 15:29:18.52                        & $-$39:50:28.3           \\
            2011 8 06.086834 & 15:29:18.54                        & $-$39:50:27.5           \\
            2011 8 06.095231 & 15:29:18.55                        & $-$39:50:26.7           \\
            2011 8 06.103634 & 15:29:18.57                        & $-$39:50:25.9           \\
            2011 8 06.112106 & 15:29:18.58                        & $-$39:50:25.1           \\
            2011 8 06.120498 & 15:29:18.60                        & $-$39:50:24.3           \\
            2011 8 06.128900 & 15:29:18.61                        & $-$39:50:23.5           \\
            2011 8 06.137431 & 15:29:18.63                        & $-$39:50:22.7           \\
            2011 8 06.145828 & 15:29:18.65                        & $-$39:50:21.9           \\
            2011 8 06.154236 & 15:29:18.67                        & $-$39:50:21.1           \\
            2011 8 06.162685 & 15:29:18.69                        & $-$39:50:20.3           \\
           \hline
        \end{tabular}
        }
        \label{astrometry2011}
     \end{table}
%
%
%
%
     \begin{table}
        \fontsize{8}{11pt}\selectfont
        \tabcolsep 0.20truecm
        \caption{Observations of 38628 Huya (2000 EB$_{173}$) over interval 2006 June 27.994178 -- 30.149815. During this observing run, 
                 Huya was in Virgo. See also Table \ref{huya}.}
        \centering
        \begin{tabular}{cccc}
           \hline
           \hline
            Date             & RA(J2000)                          & Dec(J2000)              & Filter \\
            (UT)             & ($^{\rm h}$:$^{\rm m}$:$^{\rm s}$) & (\degr:\arcmin:\arcsec) &        \\
           \hline
            2006 6 27.994178 & 14:12:43.60                        & $-$01:36:27.2           & $R$    \\
            2006 6 28.000347 & 14:12:43.58                        & $-$01:36:27.3           & $V$    \\
            2006 6 28.006551 & 14:12:43.55                        & $-$01:36:27.3           & $R$    \\
            2006 6 28.012813 & 14:12:43.55                        & $-$01:36:27.6           & $R$    \\
            2006 6 28.019028 & 14:12:43.53                        & $-$01:36:27.6           & $R$    \\
            2006 6 28.052604 & 14:12:43.47                        & $-$01:36:27.8           & $R$    \\
            2006 6 28.058831 & 14:12:43.44                        & $-$01:36:27.8           & $R$    \\
            2006 6 28.092060 & 14:12:43.37                        & $-$01:36:28.0           & $R$    \\
            2006 6 28.098079 & 14:12:43.34                        & $-$01:36:28.5           & $R$    \\
            2006 6 28.124873 & 14:12:43.30                        & $-$01:36:28.2           & $R$    \\
            2006 6 28.131030 & 14:12:43.29                        & $-$01:36:28.3           & $V$    \\
            2006 6 28.137211 & 14:12:43.29                        & $-$01:36:28.4           & $I$    \\
            2006 6 28.143403 & 14:12:43.25                        & $-$01:36:28.6           & $R$    \\
            2006 6 28.149433 & 14:12:43.23                        & $-$01:36:28.6           & $R$    \\
            2006 6 29.967211 & 14:12:39.76                        & $-$01:36:39.7           & $R$    \\
            2006 6 29.971181 & 14:12:39.73                        & $-$01:36:40.2           & $R$    \\
            2006 6 29.977211 & 14:12:39.70                        & $-$01:36:40.3           & $R$    \\
            2006 6 29.983241 & 14:12:39.69                        & $-$01:36:40.3           & $R$    \\
            2006 6 30.018438 & 14:12:39.64                        & $-$01:36:40.5           & $R$    \\
            2006 6 30.043808 & 14:12:39.58                        & $-$01:36:40.8           & $R$    \\
            2006 6 30.069086 & 14:12:39.51                        & $-$01:36:41.3           & $R$    \\
            2006 6 30.095591 & 14:12:39.46                        & $-$01:36:41.5           & $R$    \\
            2006 6 30.121945 & 14:12:39.43                        & $-$01:36:41.7           & $R$    \\
            2006 6 30.149815 & 14:12:39.40                        & $-$01:36:42.1           & $R$    \\
           \hline
        \end{tabular}
        \label{astrometryhuya2006}
     \end{table}
%
%
%
%
     \begin{table}
        \fontsize{8}{10pt}\selectfont
        \tabcolsep 0.20truecm
        \caption{Observations of 28978 Ixion (2001 KX$_{76}$) over interval 2010 May 9.267685 -- 12.396771. During this observing run, Ixion 
                 was in Ophiuchus. See also Table \ref{ix10}.}
        \centering
        \resizebox{0.40\linewidth}{0.65\linewidth}{
        \begin{tabular}{ccc}
           \hline
           \hline
            Date             & RA(J2000)                          & Dec(J2000)              \\
            (UT)             & ($^{\rm h}$:$^{\rm m}$:$^{\rm s}$) & (\degr:\arcmin:\arcsec) \\
           \hline
            2010 5 09.262489 & 17:00:21.57                        & $-$24:28:18.4           \\
            2010 5 09.267685 & 17:00:21.54                        & $-$24:28:18.2           \\
            2010 5 09.272616 & 17:00:21.53                        & $-$24:28:18.1           \\
            2010 5 09.277558 & 17:00:21.50                        & $-$24:28:18.0           \\
            2010 5 09.284294 & 17:00:21.46                        & $-$24:28:19.0           \\
            2010 5 09.292894 & 17:00:21.43                        & $-$24:28:17.7           \\
            2010 5 09.301459 & 17:00:21.39                        & $-$24:28:18.0           \\
            2010 5 09.311817 & 17:00:21.34                        & $-$24:28:18.1           \\
            2010 5 11.118172 & 17:00:12.79                        & $-$24:28:15.9           \\
            2010 5 11.128114 & 17:00:12.73                        & $-$24:28:15.8           \\
            2010 5 11.136540 & 17:00:12.68                        & $-$24:28:15.7           \\
            2010 5 11.144861 & 17:00:12.63                        & $-$24:28:15.9           \\
            2010 5 11.156586 & 17:00:12.60                        & $-$24:28:15.9           \\
            2010 5 11.166470 & 17:00:12.55                        & $-$24:28:15.9           \\
            2010 5 11.175510 & 17:00:12.50                        & $-$24:28:15.7           \\
            2010 5 11.183924 & 17:00:12.46                        & $-$24:28:15.7           \\
            2010 5 11.192327 & 17:00:12.43                        & $-$24:28:15.7           \\
            2010 5 11.200729 & 17:00:12.40                        & $-$24:28:15.6           \\
            2010 5 11.209144 & 17:00:12.34                        & $-$24:28:15.6           \\
            2010 5 11.217847 & 17:00:12.31                        & $-$24:28:15.6           \\
            2010 5 11.226551 & 17:00:12.25                        & $-$24:28:15.7           \\
            2010 5 11.234966 & 17:00:12.23                        & $-$24:28:15.9           \\
            2010 5 11.243368 & 17:00:12.19                        & $-$24:28:15.8           \\
            2010 5 11.251783 & 17:00:12.14                        & $-$24:28:15.7           \\
            2010 5 11.260197 & 17:00:12.10                        & $-$24:28:15.7           \\
            2010 5 11.269167 & 17:00:12.06                        & $-$24:28:15.6           \\
            2010 5 11.277581 & 17:00:12.00                        & $-$24:28:15.7           \\
            2010 5 11.285984 & 17:00:11.97                        & $-$24:28:15.6           \\
            2010 5 11.294398 & 17:00:11.92                        & $-$24:28:15.7           \\
            2010 5 11.302801 & 17:00:11.91                        & $-$24:28:15.9           \\
            2010 5 11.321621 & 17:00:11.81                        & $-$24:28:15.8           \\
            2010 5 11.330579 & 17:00:11.72                        & $-$24:28:15.7           \\
            2010 5 11.338993 & 17:00:11.66                        & $-$24:28:15.8           \\
            2010 5 11.347396 & 17:00:11.65                        & $-$24:28:15.8           \\
            2010 5 11.355810 & 17:00:11.62                        & $-$24:28:15.8           \\
            2010 5 11.364213 & 17:00:11.56                        & $-$24:28:15.8           \\
            2010 5 11.373021 & 17:00:11.50                        & $-$24:28:15.7           \\
            2010 5 11.381435 & 17:00:11.45                        & $-$24:28:15.7           \\
            2010 5 11.389954 & 17:00:11.42                        & $-$24:28:15.6           \\
            2010 5 11.398368 & 17:00:11.41                        & $-$24:28:15.6           \\
            2010 5 12.141204 & 17:00:07.84                        & $-$24:28:14.7           \\
            2010 5 12.149989 & 17:00:07.83                        & $-$24:28:14.6           \\
            2010 5 12.160891 & 17:00:07.78                        & $-$24:28:14.5           \\
            2010 5 12.169734 & 17:00:07.70                        & $-$24:28:14.5           \\
            2010 5 12.178704 & 17:00:07.69                        & $-$24:28:14.5           \\
            2010 5 12.187165 & 17:00:07.62                        & $-$24:28:14.5           \\
            2010 5 12.195567 & 17:00:07.61                        & $-$24:28:14.5           \\
            2010 5 12.203993 & 17:00:07.55                        & $-$24:28:14.6           \\
            2010 5 12.212396 & 17:00:07.52                        & $-$24:28:14.5           \\
            2010 5 12.220810 & 17:00:07.48                        & $-$24:28:14.5           \\
            2010 5 12.229294 & 17:00:07.44                        & $-$24:28:14.5           \\
            2010 5 12.237709 & 17:00:07.40                        & $-$24:28:14.5           \\
            2010 5 12.246123 & 17:00:07.35                        & $-$24:28:14.6           \\
            2010 5 12.254537 & 17:00:07.30                        & $-$24:28:14.5           \\
            2010 5 12.262940 & 17:00:07.27                        & $-$24:28:14.5           \\
            2010 5 12.271366 & 17:00:07.22                        & $-$24:28:14.5           \\
            2010 5 12.279861 & 17:00:07.16                        & $-$24:28:14.5           \\
            2010 5 12.288276 & 17:00:07.15                        & $-$24:28:14.4           \\
            2010 5 12.296678 & 17:00:07.10                        & $-$24:28:14.4           \\
            2010 5 12.305093 & 17:00:07.06                        & $-$24:28:14.4           \\
            2010 5 12.313507 & 17:00:06.98                        & $-$24:28:14.4           \\
            2010 5 12.322014 & 17:00:06.97                        & $-$24:28:14.4           \\
            2010 5 12.332732 & 17:00:06.90                        & $-$24:28:14.3           \\
            2010 5 12.341146 & 17:00:06.88                        & $-$24:28:14.3           \\
            2010 5 12.353172 & 17:00:06.78                        & $-$24:28:14.5           \\
            2010 5 12.363137 & 17:00:06.73                        & $-$24:28:14.5           \\
            2010 5 12.371551 & 17:00:06.69                        & $-$24:28:14.5           \\
            2010 5 12.379954 & 17:00:06.65                        & $-$24:28:14.6           \\
            2010 5 12.388357 & 17:00:06.60                        & $-$24:28:14.5           \\
            2010 5 12.396771 & 17:00:06.56                        & $-$24:28:14.5           \\
          \hline
        \end{tabular}
        }
        \label{astrometryixion2010}
     \end{table}
%
%
%
%
     \begin{table}
      \centering
          \fontsize{8}{11pt}\selectfont
          \tabcolsep 0.20truecm
          \caption{Observations of 28978 Ixion (2001 KX$_{76}$) over interval 2006 June 28.031296 -- 30.257153. During this observing run, 
                   Ixion was in Ophiuchus.}
          \begin{tabular}{ccc}
           \hline
           \hline
            Date             & RA(J2000)                          & Dec(J2000)              \\
            (UT)             & ($^{\rm h}$:$^{\rm m}$:$^{\rm s}$) & (\degr:\arcmin:\arcsec) \\
           \hline
            2006 6 28.025961 & 16:36:02.79                        & $-$22:06:19.4           \\
            2006 6 28.031296 & 16:36:02.75                        & $-$22:06:19.1           \\
            2006 6 28.065787 & 16:36:02.61                        & $-$22:06:19.1           \\
            2006 6 28.071806 & 16:36:02.58                        & $-$22:06:19.0           \\
            2006 6 28.105047 & 16:36:02.39                        & $-$22:06:18.5           \\
            2006 6 28.111077 & 16:36:02.36                        & $-$22:06:18.7           \\
            2006 6 28.156597 & 16:36:02.19                        & $-$22:06:18.3           \\
            2006 6 28.162801 & 16:36:02.13                        & $-$22:06:18.4           \\
            2006 6 28.169016 & 16:36:02.12                        & $-$22:06:18.5           \\
            2006 6 28.175232 & 16:36:02.08                        & $-$22:06:18.3           \\
            2006 6 28.181470 & 16:36:02.05                        & $-$22:06:18.4           \\
            2006 6 28.208438 & 16:36:01.94                        & $-$22:06:18.3           \\
            2006 6 28.213947 & 16:36:01.91                        & $-$22:06:18.4           \\
            2006 6 29.987709 & 16:35:53.81                        & $-$22:06:09.3           \\
            2006 6 29.991713 & 16:35:53.75                        & $-$22:06:09.2           \\
            2006 6 29.997743 & 16:35:53.74                        & $-$22:06:09.2           \\
            2006 6 30.003762 & 16:35:53.68                        & $-$22:06:09.2           \\
            2006 6 30.027107 & 16:35:53.60                        & $-$22:06:09.1           \\
            2006 6 30.052928 & 16:35:53.48                        & $-$22:06:09.0           \\
            2006 6 30.078287 & 16:35:53.37                        & $-$22:06:08.7           \\
            2006 6 30.104746 & 16:35:53.23                        & $-$22:06:08.7           \\
            2006 6 30.131412 & 16:35:53.09                        & $-$22:06:08.6           \\
            2006 6 30.159387 & 16:35:52.97                        & $-$22:06:08.7           \\
            2006 6 30.177928 & 16:35:52.90                        & $-$22:06:08.5           \\
            2006 6 30.196204 & 16:35:52.80                        & $-$22:06:08.5           \\
            2006 6 30.215648 & 16:35:52.71                        & $-$22:06:08.3           \\
            2006 6 30.236377 & 16:35:52.61                        & $-$22:06:08.0           \\
            2006 6 30.257153 & 16:35:52.52                        & $-$22:06:07.7           \\
           \hline
          \end{tabular}
          \label{astrometryixion2006}
     \end{table}
%
%
%
%
     \begin{table}
      \centering
          \fontsize{8}{11pt}\selectfont
          \tabcolsep 0.20truecm
          \caption{Observations of 90482 Orcus (2004 DW) over interval 2010 May 10.963438 -- 12.100394. During this observing run, Orcus was 
                   in Sextans. See also Table \ref{orc10}.}
          \begin{tabular}{ccc}
           \hline
           \hline
            Date             & RA(J2000)                          & Dec(J2000)              \\
            (UT)             & ($^{\rm h}$:$^{\rm m}$:$^{\rm s}$) & (\degr:\arcmin:\arcsec) \\
           \hline
            2010 5 10.963438 & 09:42:04.42                        & $-$05:48:59.6           \\
            2010 5 10.988831 & 09:42:04.51                        & $-$05:48:59.7           \\
            2010 5 10.997269 & 09:42:04.54                        & $-$05:48:59.8           \\
            2010 5 11.014688 & 09:42:04.50                        & $-$05:48:58.0           \\
            2010 5 11.031505 & 09:42:04.49                        & $-$05:48:57.9           \\
            2010 5 11.050313 & 09:42:04.48                        & $-$05:48:57.4           \\
            2010 5 11.098172 & 09:42:04.47                        & $-$05:48:56.6           \\
            2010 5 11.106609 & 09:42:04.56                        & $-$05:48:56.8           \\
            2010 5 12.057917 & 09:42:04.74                        & $-$05:48:37.0           \\
            2010 5 12.083553 & 09:42:04.78                        & $-$05:48:36.8           \\
            2010 5 12.100394 & 09:42:04.84                        & $-$05:48:35.9           \\
           \hline
          \end{tabular}
          \label{astrometryorcusmay2010}
         \end{table}
%
%
%
%
     \begin{table}
      \centering
          \fontsize{8}{11pt}\selectfont
          \tabcolsep 0.20truecm
          \caption{Observations of 90482 Orcus (2004 DW) over interval 2011 January 04.31616 -- 09.32441. During this observing run, Orcus 
                   was in Sextans. See also Table \ref{orc11}.}
          \begin{tabular}{cccc}
           \hline
           \hline
            Date             & RA(J2000)                          & Dec(J2000)              & Filter \\
            (UT)             & ($^{\rm h}$:$^{\rm m}$:$^{\rm s}$) & (\degr:\arcmin:\arcsec) &        \\
           \hline
            2011 1 04.31616  & 09:51:50.11                        & $-$07:01:27.2           & $B$    \\
            2011 1 05.31395  & 09:51:47.08                        & $-$07:01:30.9           & $V$    \\
            2011 1 05.34266  & 09:51:46.97                        & $-$07:01:31.2           & $R$    \\
            2011 1 06.27920  & 09:51:44.07                        & $-$07:01:34.3           & $R$    \\
            2011 1 06.29004  & 09:51:44.07                        & $-$07:01:34.4           & $R$    \\
            2011 1 06.33538  & 09:51:43.92                        & $-$07:01:34.4           & $R$    \\
            2011 1 06.35095  & 09:51:43.87                        & $-$07:01:34.4           & $R$    \\
            2011 1 07.33845  & 09:51:40.72                        & $-$07:01:37.1           & $R$    \\
            2011 1 09.29526  & 09:51:34.40                        & $-$07:01:41.2           & $B$    \\
            2011 1 09.32441  & 09:51:34.32                        & $-$07:01:41.2           & $R$    \\
           \hline
          \end{tabular}
          \label{astrometryorcusjanuary2011}
         \end{table}
%
%


\begin{thebibliography}{}
      \bibitem{BA05} Barucci, M.A., Belskaya, I.N., Fulchignoni, M., Birlan, M.:
                     \aj\ {\bf 130}, 1291 (2005)
      \bibitem{BB16} Batygin, K., Brown, M.E.:
                     \aj\ {\bf 151}, 22 (2016)
      \bibitem{BE10} Belskaya, I.N., Bagnulo, S., Barucci, M.A., Muinonen, K., Tozzi, G.P., Fornasier, S., Kolokolova, L.:
                     \icarus\ {\bf 210}, 472 (2010)
      \bibitem{BR14} Braga-Ribas, F., et~al.:
                     \nat\ {\bf 508}, 72 (2014)
      \bibitem{BR10} Brown, M.E., Ragozzine, D., Stansberry, J., Fraser, W.C.:
                     \aj\ {\bf 139}, 2700 (2010)
      \bibitem{BU89} Bus, S.J., Bowell, E., Harris, A.W., Hewitt, A.V.:
                     \icarus\ {\bf 77}, 223 (1989)
      \bibitem{CA09} Carraro, G.:
                     \aj\ {\bf 137}, 3089 (2009)
      \bibitem{CA06} Carraro, G., Maris, M., Bertin, D., Parisi, M.G.:
                     \aap\ {\bf 460}, L39 (2006)
      \bibitem{CL02} Clarke, D.:
                     \aap\ {\bf 386}, 763 (2002)
      \bibitem{FF14} de la Fuente Marcos, C., de la Fuente Marcos, R.:
                     \mnras\ {\bf 443}, L59 (2014)
      \bibitem{FM15} de la Fuente Marcos, C., de la Fuente Marcos, R., Aarseth, S.J.:
                     \mnras\ {\bf 446}, 1867 (2015)
      \bibitem{DB07} Di Sisto, R.P., Brunini, A.:
                     \icarus\ {\bf 190}, 224 (2007)
      \bibitem{DU14} Duffard, R., et~al.:
                     \aap\ {\bf 568}, A79 (2014)
      \bibitem{DW83} Dworetsky, M.M.:
                     \mnras\ {\bf 203}, 917 (1983)
      \bibitem{FO13} Fornasier, S., et~al.:
                     \aap\ {\bf 555}, A15 (2013)
      \bibitem{FO14} Fornasier, S., et~al.:
                     \aap\ {\bf 568}, L11 (2014)
      \bibitem{FU08} Fulchignoni, M., Belskaya, I.N., Barucci, M.A., de Sanctis, M.C., Doressoundiram, A.:
                     In: Barucci, M.A., Boehnhardt, H., Cruikshank, D.P., Morbidelli, A., (eds.) The Solar System Beyond Neptune.
                     p.\ 181. University of Arizona Press, Tucson (2008)
      \bibitem{GA09} Galiazzo, M.:
                     Masters Thesis, Universit\`a degli Studi di Padova, Italy (2009)
      \bibitem{GA15} Galiazzo, M., Carruba, V., Wiegert, P.:
                     IAU General Assembly, Meeting \#29, id. 2252424 (2015)
      \bibitem{GL08} Gladman, B., Marsden, B.G., Vanlaerhoven, C.:
                     In: Barucci, M.A., Boehnhardt, H., Cruikshank, D.P., Morbidelli, A., (eds.) The Solar System Beyond Neptune.
                     p.\ 43. University of Arizona Press, Tucson (2008)
      \bibitem{GL12} Gladman, B., et~al.:
                     \aj\ {\bf 144}, 23 (2012)
      \bibitem{HD02} Hainaut, O.R., Delsanti, A.C.:
                     \aap\ {\bf 389}, 641 (2002)
      \bibitem{HA06} Hamuy, M., et~al.:
                     \pasp\ {\bf 118}, 2 (2006)
      \bibitem{JE09} Jewitt, D.:
                     \aj\ {\bf 137}, 4296 (2009)
      \bibitem{JD84} Jewitt, D., Danielson, G.E.: 
                     \icarus\ {\bf 60}, 435 (1984)
      \bibitem{JL93} Jewitt, D., Luu, J.:
                     \nat\ {\bf 362}, 730 (1993)
      \bibitem{LL06} Lacerda, P., Luu, J.:
                     \aj\ {\bf 131}, 2314 (2006)
      \bibitem{LK65} Lafler, J., Kinman, T.D.:
                     \apjs\ {\bf 11}, 216 (1965)
      \bibitem{LA92} Landolt, A.U.:
                     \aj\ {\bf 104}, 372 (1992)
      \bibitem{LA10} Lang, D., Hogg, D.W., Mierle, K., Blanton, M., Roweis, S.:
                     \aj\ {\bf 139}, 1782 (2010)
      \bibitem{LE13} Lellouch, E., et~al.:
                     \aap\ {\bf 557}, A60 (2013)
      \bibitem{LD97} Levison, H., Duncan, M.:
                     \icarus\ {\bf 127}, 13 (1997)
      \bibitem{LE01} Levison, H., Dones, L., Duncan, M.:
                     \aj\ {\bf 121}, 2253 (2001)
      \bibitem{LR07} Lorin, O., Rousselot, P.:
                     \mnras\ {\bf 376}, 881 (2007)
      \bibitem{LJ90} Luu, J.X., Jewitt, D.C.:
                     \aj\ {\bf 100}, 913 (1990)
      \bibitem{MB93} Marcialis, R.L., Buratti, B.J.:
                     \icarus\ {\bf 104}, 234 (1993)
      \bibitem{NA10} Naoz, S., Perets, H.B., Ragozzine, D.:
                     \apj\ {\bf 719}, 1775 (2010)
      \bibitem{OR03} Ortiz, J.L., Guti\'errez, P.J., Casanova, V., Sota, A.:
                     \aap\ {\bf 407}, 1149 (2003)
      \bibitem{OR06} Ortiz, J.L., Guti\'errez, P.J., Santos-Sanz, P., Casanova, V., Sota, A.:
                     \aap\ {\bf 447}, 1131 (2006)
      \bibitem{OE11} Ortiz, J.L., et~al.:
                     \aap\ {\bf 525}, 31 (2011)
      \bibitem{OR15} Ortiz, J.L., et~al.:
                     \aap\ {\bf 576}, A18 (2015)
      \bibitem{PA11} Parker, A.H.:
                     Ph.D. Thesis, University of Victoria, Canada (2011)
      \bibitem{PK11} Parker, A.H., Kavelaars, J.J., Petit, J.-M., Jones, L., Gladman, B., Parker, J.:
                     \apj\ {\bf 743}, 1 (2011)
      \bibitem{PR07} Press, W.H., Teukolsky, S.A., Vetterling, W.T., Flannery, B.P.:
                     Numerical Recipes: The Art of Scientific Computing, 3rd Edition.
                     Cambridge University Press, Cambridge (2007)
      \bibitem{RA07} Rabinowitz, D.L., Schaefer, B.E., Tourtellotte, S.W.:
                     \aj\ {\bf 133}, 26 (2007)
      \bibitem{RT05} Romanishin, W., Tegler, S.C.:
                     \icarus\ {\bf 179}, 523 (2005)
      \bibitem{RP10} Rousselot, P., Petit, J.:
                     AAS, DPS meeting \#42, \#40.19 (2010)
      \bibitem{RU15} Ruprecht, J.D., et~al.:
                     \icarus\ {\bf 252}, 271 (2015)
      \bibitem{SH07} Sheppard, S.S.:
                     \aj\ {\bf 134}, 787 (2007)
      \bibitem{GL08} Sheppard, S.S., Lacerda, P., Ortiz, J.L.:
                     In: Barucci, M.A., Boehnhardt, H., Cruikshank, D.P., Morbidelli, A., (eds.) The Solar System Beyond Neptune.
                     p.\ 129. University of Arizona Press, Tucson (2008)
      \bibitem{ST87} Stetson, P.B.:
                     \pasp\ {\bf 99}, 191 (1987)
      \bibitem{TH10} Thirouin, A., Ortiz, J.L., Duffard, R., Santos-Sanz, P., Aceituno, F.J., Morales, N.:
                     \aap\ {\bf 522}, A93 (2010)
      \bibitem{TH14} Thirouin, A., Noll, K.S., Ortiz, J.L., Morales, N.:
                     \aap\ {\bf 563}, A3 (2014)
      \bibitem{TS14} Trujillo, C.A., Sheppard, S.S.:
                     \nat\ {\bf 507}, 471 (2014)
     \bibitem{iWJ13} Wall, J.V., Jenkins, C.R.:
                     Practical Statistics for Astronomers, 2nd Edition.
                     Cambridge University Press, Cambridge (2012)
      \bibitem{WA09} Walsh, K.J.:
                     Earth, Moon, and Planets\ {\bf 105}, 193 (2009)
   \end{thebibliography}
\end{document}